\documentclass[journal=jcisd8,manuscript=article]{achemso}
\setkeys{acs}{etalmode=truncate,maxauthors=10}
\SectionNumbersOn
\usepackage[utf8]{inputenc}
\usepackage[T1]{fontenc}
\usepackage{graphicx}
\usepackage{dcolumn}
\usepackage{bm}
\usepackage{siunitx}
\usepackage{hyperref}
\usepackage{rotating}
\usepackage{xcolor}
\usepackage{amsmath}
\usepackage{amsfonts}
\usepackage{mathtools}
\usepackage[version=4]{mhchem}
\usepackage{booktabs}
\usepackage{multirow}
\usepackage{makecell}

\graphicspath{{figures/}}

\newcommand{\changed}[1]{{#1}}
\newcommand{\changedagain}[1]{{#1}}

\newcommand{\onlinecite}[1]{\citenum{#1}}
\renewcommand{\vec}[1]{\mathrm{\mathbf{#1}}}
\renewcommand{\Re}{\mathbb R}
\DeclareMathOperator{\Neigh}{Neigh}

\usepackage{xspace}
\makeatletter
\def\ie{\textit{i.e.}\@\xspace}
\def\eg{\textit{e.g.}\@\xspace}
\def\vs{\textit{vs.}\@\xspace}
\DeclareRobustCommand\onedot{\futurelet\@let@token\@onedot}
\def\@onedot{\ifx\@let@token.\else.\null\fi\xspace}

\def\etal{\textit{et al}\onedot}
\makeatother

\newcommand{\gdb}{GDB7-22-TS\xspace}
\newcommand{\proparg}{Proparg-21-TS\xspace}
\newcommand{\cyclo}{Cyclo-23-TS\xspace}

\newcommand{\extref}[1]{\ref*{S-#1}}
\newcommand{\LCMD}{Laboratory for Computational Molecular Design, Institute of Chemical Sciences and Engineering,
\'{E}cole Polytechnique F\'{e}d\'{e}rale de Lausanne, 1015 Lausanne, Switzerland}
\newcommand{\LAS}{Learning \& Adaptive Systems Group, Department of Computer Science, ETH Zurich, 8092 Zurich, Switzerland}
\newcommand{\NCCRcat}{National Center for Competence in Research -- Catalysis (NCCR-Catalysis), \'{E}cole Polytechnique F\'{e}d\'{e}rale de Lausanne, 1015 Lausanne, Switzerland}

\author{Puck van Gerwen} % 0000-0002-7992-5529
\altaffiliation{These authors contributed equally to this work.}
\affiliation{\LCMD}
\alsoaffiliation{\NCCRcat}
\author{Ksenia R. Briling} % 0000-0003-2250-9898
\altaffiliation{These authors contributed equally to this work.}
\affiliation{\LCMD}
\author{Charlotte Bunne} % 0000-0003-1431-103X
\affiliation{\NCCRcat}
\alsoaffiliation{\LAS}
\author{Vignesh Ram Somnath} % 0000-0003-3894-646X
\affiliation{\NCCRcat}
\alsoaffiliation{\LAS}
\author{Ruben Laplaza} % 0000-0001-6315-4398
\affiliation{\LCMD}
\alsoaffiliation{\NCCRcat}
\author{Andreas Krause} % 0000-0001-7260-9673
\affiliation{\NCCRcat}
\alsoaffiliation{\LAS}
\author{Clemence Corminboeuf} % 0000-0001-7993-2879
\email{clemence.corminboeuf@epfl.ch}
\affiliation{\LCMD}
\alsoaffiliation{\NCCRcat}

\title{\changed{3DReact: Geometric deep learning} \texorpdfstring{\\}{} for chemical reactions}

\keywords{machine learning, equivariant neural networks, geometric deep learning, activation energies, chemical reactions}

\begin{document}

\begin{abstract}
\changed{Geometric deep learning models,} which incorporate the relevant molecular symmetries within the neural network
architecture, have considerably improved the accuracy and
data efficiency of predictions of molecular properties. Building on this
success, we introduce
\changed{\textsc{3DReact}, a geometric deep learning model} to predict reaction properties
from three-dimensional structures of reactants and products. \changed{We demonstrate that the invariant version of the model is
sufficient for existing reaction datasets.} We illustrate its competitive performance on the prediction of activation barriers
on the \gdb, \cyclo and \proparg
datasets
in different atom-mapping regimes.
We show that, compared to existing models for reaction
property prediction,
\changed{\textsc{3DReact} offers a flexible framework that exploits atom-mapping information, if available, as well as
geometries of reactants and products (in an invariant or equivariant fashion). Accordingly, it performs systematically well
across different datasets, atom-mapping regimes, as well as both interpolation and extrapolation tasks.}
\end{abstract}

%%%%%%%%%%%%%%%%%%%%%%%%%%%%%%%%%%%%%%%%%%%%%%%%%%%%%%%%%%%%%%%%%%%%%%%%%%%%%%%%%%%%%%%%%%%%%%%%%%%%
\section{Introduction}
\label{sec:intro}

Physics-inspired representations that take as input the three-dimensional
structure\cite{behler2007, ruppfastandaccurate2012, soapbartok2013,
bagofbonds2015, huo2017unified, fchl18, fchl19, amons2020, drautz2019atomic,
dusson2022atomic, grisafi2019incorporating, grisafi2021multi, nigam2020recursive} (as well as, in some
cases, electronic structure\cite{spahm_2022, spahm_2023, karandashev_2022, LG2023}) of
molecules and transform it into a fixed-length vector, while respecting known physical laws, have a rich
history in molecular property prediction.\cite{ruppfastandaccurate2012,
amons2020, bagofbonds2015,  fchl18, fchl19, nigam2020recursive, li2015molecular,
chmiela2017machine, chmiela2018towards,
huo2017unified, behler2007,
behler2017first, soapbartok2013, drautz2019atomic,
dusson2022atomic, smith2018less, bereau2015transferable,grisafi2018symmetry, wilkins2019accurate,
grisafi2021multi, montavon2013machine, mazouin2022selected, brockherde2017bypassing, grisafi2018transferable, fabrizio2019electron}
Common desiderata\cite{musil_2021_review, langer_2022_review, huang_2021_review, Kulik_2022_review}
for high-performing representations are (i) smoothness, (ii)
encoding of the appropriate symmetries to permutations, rotations and
translations,\cite{grisafi2018symmetry, glielmo2017accurate} (iii) completeness
and (iv) additivity to allow for extrapolation to larger systems. Such fingerprints,\cite{ruppfastandaccurate2012, bagofbonds2015,
soapbartok2013,grisafi2018symmetry, fchl18, fchl19, amons2020, huo2017unified,
grisafi2019incorporating, grisafi2021multi, nigam2020recursive}
being rooted in fundamental principles, are designed to be property-independent: a single
representation can be constructed for a molecule to predict any
quantum-chemical target. This is analogous to the
molecular Hamiltonian, which specifies the energy and all other properties of a system as a function
of atoms' types and positions in three-dimensional space (assuming the molecules are charge neutral and singlets).
These representations are typically used in combination with kernel models due to their data efficiency,
ability to deal with high-dimensional feature vectors, and interpretability of the similarity
kernel.\cite{ruppfastandaccurate2012, soapbartok2013, bagofbonds2015,huo2017unified, fchl18, fchl19, amons2020,
drautz2019atomic, dusson2022atomic,grisafi2021multi, nigam2020recursive, musil_2021_review,
langer_2022_review,huang_2021_review} Early works showed that combining such representations\cite{ruppfastandaccurate2012,
bagofbonds2015, fchl18, amons2020, vangerwen_2022} with simple feed-forward neural networks instead of kernel models did not
necessarily led to better performance.\cite{faber2017prediction, digdisc}

More recently, end-to-end neural networks have been proposed that learn the representation
as part of the (supervised) training process,\cite{schutt2017schnet,
unke2019physnet, gasteiger2020directional, gilmer2017neural, batzner_e3nn_2022,
gasteiger2021gemnet, haghighatlari2022newtonnet, qiao2020orbnet,
thomas2018tensor, townshend2020geometric, anderson2019cormorant,
pmlr-v139-satorras21a,
christensen2021orbnet,
pmlr-v139-schutt21a,
unke2021spookynet,
zhang2020efficient, nguyen2022predicting,
mace_2022,
liao2022equiformer,
fuchs2020se,
Simeon_2023, Corso_2024, Duval_2023} based on similar principles to the aforementioned physics-inspired representations:
they take as input a three-dimensional structure, as well as
in some cases charge and spin information.\cite{qiao2020orbnet, christensen2021orbnet, pmlr-v139-schutt21a, unke2021spookynet}
\changed{The network may be \textit{invariant} or \textit{equivariant} to rotations and translations of the input molecules.
The former is typically achieved by operating on distances between atoms,\cite{schutt2017schnet, unke2019physnet, gilmer2017neural}
and the latter by operating on relative position vectors and angular information
processed by rotationally-equivariant convolutional layers.\cite{qiao2020orbnet, anderson2019cormorant,
pmlr-v139-satorras21a, musaelian2023learning, Simeon_2023, haghighatlari2022newtonnet, batzner_e3nn_2022,
gasteiger2020directional, gasteiger2021gemnet, townshend2020geometric, zhang2020efficient,
nguyen2022predicting, mace_2022, liao2022equiformer, fuchs2020se}
Equivariant models are naturally suited to predict vectorial\cite{gasteiger2021gemnet, anderson2019cormorant,
musaelian2023learning, batzner_e3nn_2022, Simeon_2023, townshend2020geometric, haghighatlari2022newtonnet}
or higher order tensorial\cite{Simeon_2023, pmlr-v139-schutt21a, zhang2020efficient, nguyen2022predicting, Wen_2024} properties.
They have also been demonstrated to exhibit improved data efficiency and generalization capabilities
compared to their invariant counterparts on predictions of scalar properties,\cite{batzner_e3nn_2022}
\changedagain{albeit at a higher computational cost.} Nevertheless, given an expressive enough architecture
(\ie, using higher-order messages\cite{gasteiger2020directional, musaelian2023learning, batatia2022design_botnet,
liu2022spherical, mace_2022, kondor2018nbody, multi_ace_2022} and/or enough convolutional
layers\cite{batzner_e3nn_2022, pmlr-v139-schutt21a, mace_2022}),
invariant models are sufficient for many property prediction tasks.\cite{mace_2022}
}

Despite these advances for molecular property prediction, the prediction of
computed \emph{reaction properties} (principally, reaction
barriers\cite{Atwell2022, vangerwen_2022, digdisc, Gallarati2021,
grambow_deep_learning, Heid2021, spiekermann_predict, isayev_delta2,
Heinen2021, Singh2019, Choi2018, farrar_semiemp_barriers,
dihydrogen_friederich, Migliaro2020,lewis2023reformulating, Vadaddi_2024, Ramos_2024}) is still in its
infancy.\cite{Schwaller_Review_2022} Machine learning approaches span from utilizing simple two-dimensional fingerprints of
reaction components\cite{Rogers2010, Probst2022} (reactants and products) to physical-organic descriptors\cite{Doyle2018,
Zuranski2021, Choi2018, Denmark2019, Jorner2021, Singh2019, reid2019holistic, gensch2022comprehensive, santiago2018predictive,
jorner_review, paton_physorg_2021, sigman_doyle_2021, lewis2023reformulating, stuyver_model_2023, stuyver_qmgnn, Vargas_2024,
Ramos_2024}, or electronic structure-inspired features\cite{Vijay_2024}, to transformer models\cite{Devlin2018,
schwaller_transfomer} adapted for regression,\cite{Schwaller_Yields_2021} and 2D graph-based approaches\cite{Heid2021,
heid2023chemprop, grambow_deep_learning, Vadaddi_2024}.
The latter, particularly the \textsc{ChemProp} model,\cite{Heid2021, heid2023chemprop}
are often best-in-class in predicting reaction properties.\cite{heid2023chemprop}
It has been shown\cite{digdisc} that these models achieve their impressive performance by exploiting atom-mapping
information,\cite{chen2013automatic, preciat2017comparative,
jaworski2019automatic, schwaller2021extraction} which provide information analogous to the reaction mechanism.

Another category of reaction fingerprints arises from discretization of physically-inspired
functions\cite{ruppfastandaccurate2012, soapbartok2013, bagofbonds2015, huo2017unified, fchl18, fchl19, amons2020,
drautz2019atomic, dusson2022atomic, grisafi2021multi, nigam2020recursive} constructed using a cheap estimate of the transition
state (TS) structure\cite{isayev_delta2} or rather the structures of the reaction components\cite{vangerwen_2022,
Gallarati2021, Heinen2021} The SLATM$_d$ representation\cite{Gallarati2021, vangerwen_2022} in particular has been
shown\cite{digdisc} to yield accurate predictions of reaction barriers, particularly for datasets\cite{Doney2016,
Gallarati2021} relying on subtle changes in the geometry of reactants and/or products.
End-to-end models based on three-dimensional structures of reactants and products have also recently emerged.\cite{Vijay_2024,
spiekermann_predict, Nehil_Puleo_2024}
\changed{In a different vein, several works\cite{duan2023accurate, Zhang_2021, Pattanaik_2020, Makos_2021, Kim_2023, Choi_2023}
aim to directly predict the TS structure, which together with the reactant structure gives the reaction barrier.
These approaches lie outside the scope of the property-prediction focus here.}

\changed{Due to the diversity of challenges posed by different reaction datasets,
neither atom-mapping-based models nor 3D-geometry-based models
achieve consistently better performance on reaction property prediction tasks.\cite{digdisc}
To date, no model has been proposed that can incorporate both chemical (atom-maps) and physical (geometry) priors.
To address this gap,} we introduce
\textsc{3DReact},
a geometric deep learning model that encodes both the three-dimensional structures of reactants and products
as well as atom-mapping information or proxies thereof
to predict properties of chemical reactions (showcased here for activation energies).

We demonstrate the performance of
\textsc{3DReact}
on three datasets of reaction barriers:
\gdb,\cite{gdb7-22-ts} \cyclo,\cite{coley_dataset} and \proparg.\cite{Doney2016, Gallarati2021}
As discussed in previous works,\cite{digdisc}
these datasets present a myriad of challenges for ML models,
from the dependence on chemical information\cite{gdb7-22-ts}
to the distinction of subtle changes in configurations.\cite{Doney2016, Gallarati2021}
We show that, compared to state-of-the-art models for reaction property prediction,\cite{heid2023chemprop, vangerwen_2022}
\changed{\textsc{3DReact} offers accurate and reliable performance across different datasets
as well as atom-mapping regimes, reduced dependence on the quality of three-dimensional geometries,
and stable extrapolation behavior.}

%%%%%%%%%%%%%%%%%%%%%%%%%%%%%%%%%%%%%%%%%%%%%%%%%%%%%%%%%%%%%%%%%%%%%%%%%%%%%%%%%%%%%%%%%%%%%%%%%%%%
\section{Architecture}
\label{sec:arch}

\begin{figure}[bth]
\centering
\includegraphics[width=1\linewidth]{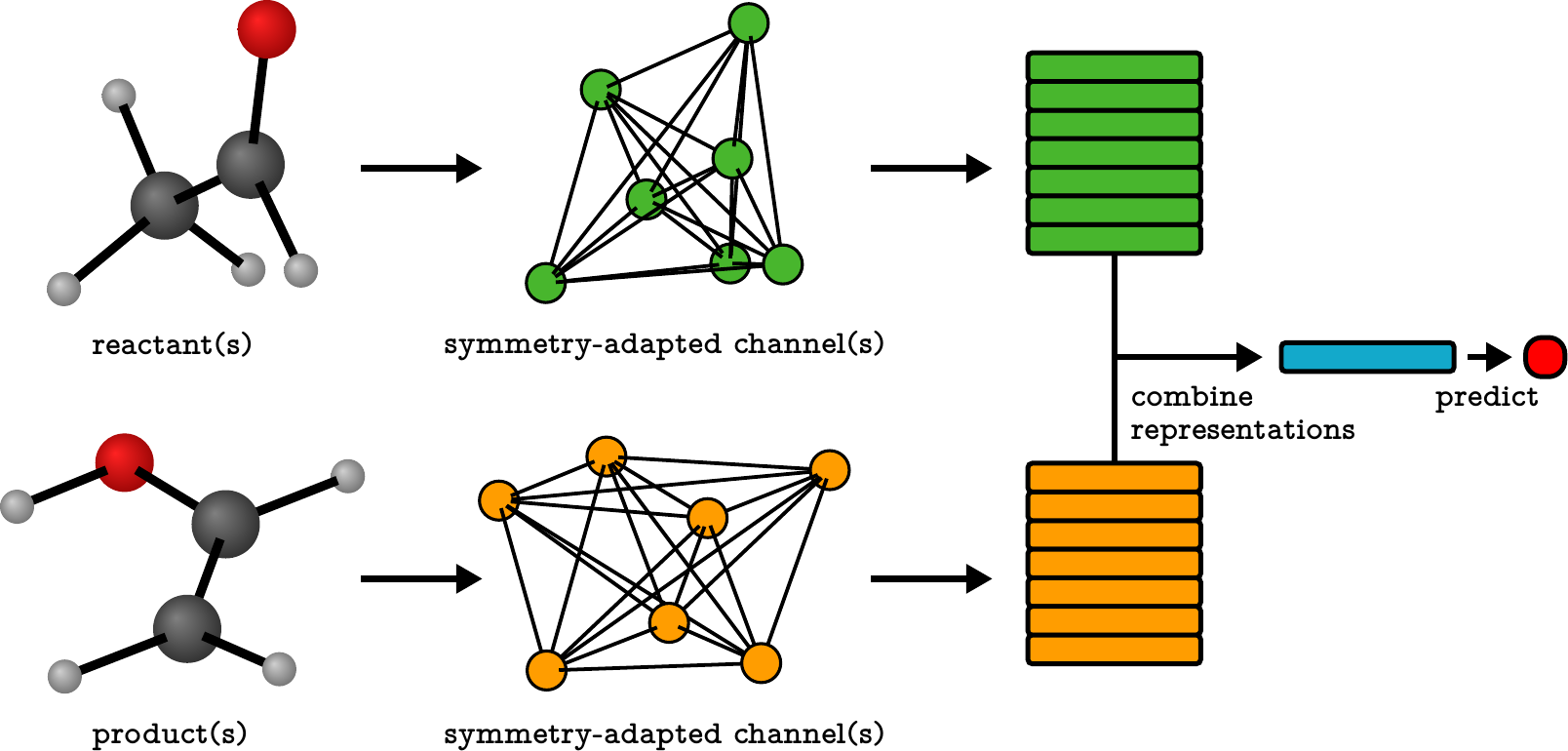}
\caption{Architecture of
\textsc{3DReact}.
Molecules pass through independent
\changed{symmetry-adapted (invariant or equivariant)} channels (green and orange).
These are combined to yield a reaction representation (blue) which is used to predict a reaction property,
such as the activation energy (red dot).}
\label{fig:EquiReact}
\end{figure}

\textsc{3DReact}
is built from $\mathrm{O}(3)$-equivariant convolutional networks over point
clouds as implemented in \texttt{e3nn}.\cite{geiger2020euclidean}
Specifically, we use the tensor field network architecture\cite{thomas2018tensor}
for molecular components as in Corso \etal.~\cite{corso2023diffdock}
\changed{While the architecture is equivariant by default, it can easily be made invariant (\textit{vide infra}).}
The geometries of molecules constituting reactants and products of each reaction are
passed through separate channels,
detailed in Section~\ref{sec:arch:molecular_channels}.
They are then combined to eventually predict a reaction property,
such as the activation energy, as detailed in Section~\ref{sec:arch:reaction}.
The overall architecture is summarized in Figure~\ref{fig:EquiReact}.

%%%%%%%%%%%%%%%%%%%%%%%%%%%%%%%%%%%%%%%%%%%%%%%%%%%%%%%%%%%%%%%%%%%%%%%%%%%%%%%%%%%%%%%%%%%%%%%%%%%%
\subsection{Symmetry-adapted molecular channels}
\label{sec:arch:molecular_channels}

A molecule with $N_{\mathrm{at}}$ atoms is represented as a distance-based graph where nodes describe atoms and edges
describe bonds. Instead of explicitly using connectivity information, the
``bonds'' of atom $a$ are formed with all the neighboring $\Neigh(a)$ atoms within the
cutoff $r_{\max}$.
\changed{Initial scalar bond (edge) features $\{\vec e^{(0)}_{ab}\}$ between atoms $a$ and $b$,
as well as spherical harmonics filters $\{\vec z_{ab}\}$, are computed from internal coordinates,
as detailed in equations~\extref{eq:initial-edge-features-start}--\extref{eq:initial-edge-features-end}.}
The atom (node) features $\{\vec x^{(0)}_a\}$ are initialized with $n_f{=}16$~cheminformatics descriptors
computed with \texttt{RDKit}.\cite{rdkit2023} These include
atomic number,
chirality tag (unspecified, tetrahedral, or other, including octahedral, square planar, allene-type),
number of directly-bonded neighbors, number of rings,
implicit valence, formal charge, number of attached hydrogens, number of unpaired
electrons, hybridization, aromaticity,
and presence in rings of specified sizes from $3$ to $7$.
This choice is inspired by \textsc{EquiBind}\cite{stark2022equibind} and \textsc{DiffDock}\cite{corso2023diffdock}
and is in line with the improved~\cite{spiekermann_predict} features used for 2D-based methods.

The initial node and edge features pass through embeddings to give $\{\vec x^{(1)}_a\}$ and $\{\vec{e}_{ab}\}$ respectively,
the former are then updated by $n_{\mathrm{conv}} \in \{2,3\}$ equivariant convolutional layers.
Each layer is a fully-connected weighted tensor product, as defined in \texttt{e3nn}\cite{geiger2020euclidean}.
\changed{Equations~\extref{eq:convs-start}--\extref{eq:convs-end} describe the equivariant operations
performed by the network (see Section~\extref{sec:molecular_channels} for mathematical details).
The network with equivariant molecular components as described is referred to as \textsc{EquiReact},
where its invariant counterpart \textsc{InReact} uses only the $\ell=0$ (scalar) spherical harmonics
to construct the convolution filters.
The output of the molecular channels is} the local molecular representation $\vec X\in \Re^{N_{\mathrm{at}} \times D}$
corresponding to $N_{\mathrm{at}}$ atoms associated with $D$ features.
Depending on the \texttt{sum\_mode} hyperparameter,
it is constructed either from the node features (\texttt{node} mode)
or both node and edge features (\texttt{both} mode).

Inspired by the \textsc{ChemProp} model,\cite{Heid2021, heid2023chemprop}
we added an option to exclude hydrogen atoms as nodes when constructing the graph.
The only information about hydrogens is then contained in the initial edge features of heavy atoms.

%%%%%%%%%%%%%%%%%%%%%%%%%%%%%%%%%%%%%%%%%%%%%%%%%%%%%%%%%%%%%%%%%%%%%%%%%%%%%%%%%%%%%%%%%%%%%%%%%%%%
\subsection{Combining molecules for reactions}
\label{sec:arch:reaction}

Once atom-wise molecular representations $\vec X$ are learned for reactant and
product molecules, they must be combined to form a reaction representation
$\vec X_{\mathrm{rxn}}$.

\begin{figure}[ht!]
\centering
\includegraphics[width=0.9\textwidth]{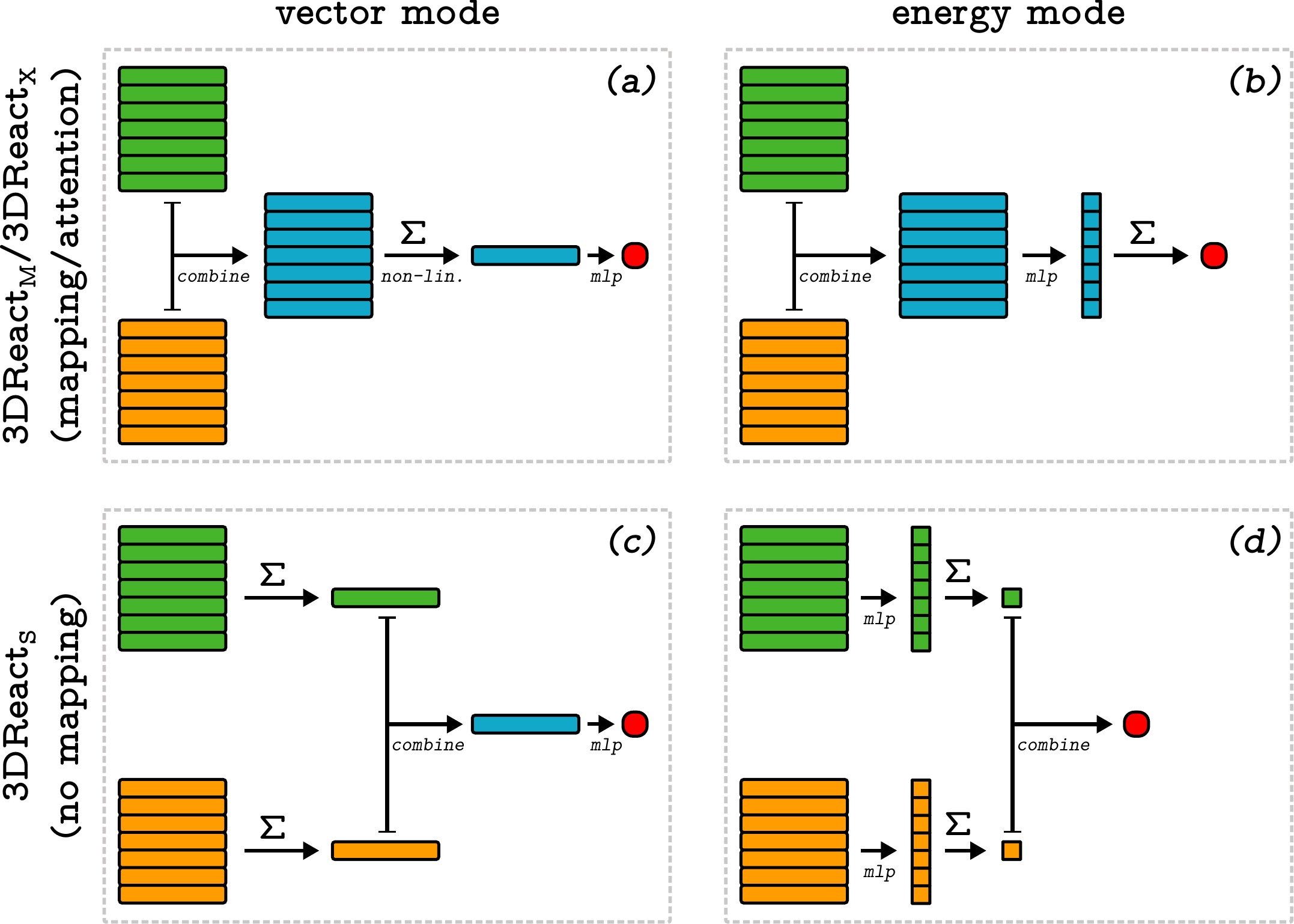}
\caption{Scheme illustrating how the reactant (green) and product (orange) representations
are combined to form a reaction representation (blue)
and eventually predict the target property (red dot) using a multilayer perceptron (mlp).
$\sum$ refers to the summation over atom-wise environments.
Oblong rectangles and squares represent vectors and scalars, respectively.
}
\label{fig:combine}
\end{figure}

For certain datasets, atom-mapping information is available, which correlates
individual atoms in reactant molecules to individual atoms in product
molecules according to the reaction mechanism. In this setting, the representations $\vec X_{\mathrm{reactant}}$ and
$\vec X_{\mathrm{product}}$ are re-ordered such that the local representation vectors
correspond to the same atom in reactants and products.
Depending on the \texttt{combine\_mode} hyperparameter, either a
difference is taken between products' and reactants' atom representations, or
they are summed, averaged, or passed through a multilayer perceptron (MLP).
Thus, the local reaction representation $\vec X_{\mathrm{rxn}}$ consists
of vectors reflecting how the environment changes in the reaction for each atom.
We will refer to this variant of the model, which uses atom-mapping information, as \textsc{3DReact}$_M$.
While the current model is unable to treat unbalanced reactions
(where there are additional atoms on the left- or right-hand side of the reaction equation),
its modification in the spirit of \textsc{ChemProp}\cite{Heid2021, heid2023chemprop}
is straightforward.

With the reaction representation at hand,
predictions are made in the so-called \texttt{vector} or \texttt{energy} modes.
In \texttt{vector} mode,
the atomic vectors constituting the reaction representation $\vec X_{\mathrm{rxn}}$
are initially passed through an MLP to introduce nonlinearity and then
summed up to form a global reaction representation vector $\vec {\bar X}_{\mathrm{rxn}}$.
The target is then learned using an MLP.
This model pipeline is illustrated in Figure~\ref{fig:combine}a.
In \texttt{energy} mode, on the other hand, the local reaction representations
are used to learn atomic contributions to the target (Figure~\ref{fig:combine}b).
While performing worse in general,
in some cases this mode yields the best predictions
(see Section~\ref{sec:results:performance}).

Atom-mapping provides \textit{static} information, analogous to a reaction mechanism,
to link atoms in reactants to atoms in products.
\changed{While highly informative, and thought to be critical to the performance of 2D-graph-based models,
\cite{Heid2021, heid2023chemprop, spiekermann_predict,grambow_deep_learning, Vadaddi_2024}
accurate atom-maps are not available for all reaction datasets.\cite{digdisc, chen2013automatic, preciat2017comparative}}
\changed{To circumvent the need for atom-mapping, but mimic its role in exchanging information between reactants and products,
other approaches} \textit{dynamically} (\ie, in a learnable fashion) exchange information between molecular representations.
For example, \textsc{RXNMapper}\cite{schwaller2021extraction} is a neural network
that learns atom-mappings within the larger self-supervised task
of predicting the randomly masked parts in a reaction sequence,
using one head of a multi-head transformer architecture.
\textsc{EquiBind},\cite{stark2022equibind} a neural network that predicts
the rotation and translation of a ligand to a protein,
contains a cross-attention module between ligand and receptor.
The latter inspires our surrogate for atom-mapping: \textsc{3DReact}$_X$ also uses
cross-attention between reactants and products to link their atom indices (Section~\extref{sec:cross}).
The re-ordered representations of reactants and products are combined
as for the case of atom-mapped reactions
(Figures~\ref{fig:combine}a and \ref{fig:combine}b).
We note that other algorithms could also have been used to exchange information between reactants and products,
for example in the form of message passing or equivariant attention.\cite{liao2022equiformer, ganea2022independent}

\textsc{3DReact} also has a simple ``no mapping'' variant,
called \textsc{3DReact}$_S$,
which does not rely on atom-mapping, nor a surrogate cross-attention module.
In \texttt{vector} mode (Figure~\ref{fig:combine}c),
the atomic components of molecular representations $\vec X_{\mathrm{reactant}}$
and $\vec X_{\mathrm{product}}$ are summed up
to obtain global vectors $\vec {\bar X}_{\mathrm{reactant}}$
and $\vec {\bar X}_{\mathrm{product}}$, respectively.
Then they are combined, according to the \texttt{combine\_mode} parameter,
to form a reaction vector $\vec {\bar X}_{\mathrm{rxn}}$ which is used to
learn the target with an MLP.
In \texttt{energy} mode (Figure~\ref{fig:combine}d)
individual atomic representations are used to learn their
contributions to the quasi-molecular energies of reactants and products,
which are later combined
(according to the \texttt{combine\_mode} parameter)
to predict the target.
In most cases, this simpler model out-performs \textsc{3DReact}$_X$ (\textit{vide infra}).

%%%%%%%%%%%%%%%%%%%%%%%%%%%%%%%%%%%%%%%%%%%%%%%%%%%%%%%%%%%%%%%%%%%%%%%%%%%%%%%%%%%%%%%%%%%%%%%%%%%%
\section{Results and Discussion}
\label{sec:results}
The performance of \textsc{3DReact} is reported for three diverse datasets
(the \gdb,\cite{gdb7-22-ts} \cyclo\cite{coley_dataset} and \proparg\cite{Doney2016, Gallarati2021})
using both random and \changed{extrapolative splits. For details on the datasets,
refer to Section~\ref{sec:methods:datasets}. For details on the extrapolation splits, see Section~\ref{sec:methods:splits}.}

Models are run in three atom-mapping regimes:
(i)~with high-quality maps (``True'')
derived from the TS structures
or heuristic rules;\cite{coley_dataset, Grambow2020, gdb7-22-ts, Heid2021, jaworski2019automatic}
(ii)~with atom-maps obtained using the open-source \textsc{RXNMapper}\cite{schwaller2021extraction} (``RXNMapper'');
and
(iii)~without any atom-mapping information at all (``None'').
As discussed in recent work,\cite{reply_comment, digdisc}
previously developed graph-based models for reaction property prediction\cite{Heid2021, spiekermann_predict,
grambow_deep_learning, stuyver_model_2023, stuyver_qmgnn}
including \textsc{ChemProp}\cite{Heid2021, heid2023chemprop}
reported prediction errors only in the ``True'' atom-mapping regime.
The ``RXNMapper'' regime is important for cases where the reaction mechanism is not known
and atom-mapping using heuristic rules is impossible.
The ``None'' regime is critical for all chemistry that falls outside the realm of organic chemistry
captured in the patents\cite{lowe2012extraction}
\mbox{that \textsc{RXNMapper}\cite{schwaller2021extraction} is trained on.}

The atom-mapping-based model \textsc{3DReact}$_M$
is used in the ``True'' and ``RXNMapper'' regimes.
In the ``None'' regime, \textsc{3DReact}$_X$ and \textsc{3DReact}$_S$ were tested.
\textsc{3DReact}$_S$ consistently outperformed \textsc{3DReact}$_X$,
so we include only \textsc{3DReact}$_S$ and refer the reader to Section~\extref{sec:cross} for their comparison.

\subsection{\changed{Equivariance \texorpdfstring{\vs}{vs.} invariance}}
\label{sec:results:eq_vs_inv}

\begin{figure}[tbh!]
\centering
\includegraphics[width=0.98\linewidth]{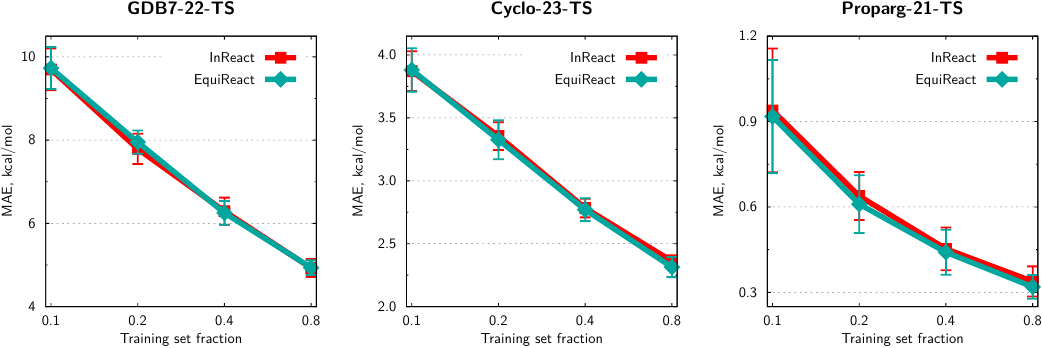}
\caption{Learning curves for \textsc{InReact} and \textsc{EquiReact} in the ``True'' atom-mapping regime.
Each point shows mean absolute error (MAE), averaged over 10~folds of 80/10/10 splits
(for training set fraction $< 0.8$, the corresponding subset of the ``full'' training set is used),
and error bars indicate standard deviations across folds.
}
\label{fig:learning_curves}
\end{figure}

\changed{Table~\ref{tab:model_performance_equireact} compares the relative performance of the invariant (\textsc{InReact})
and the equivariant (\textsc{EquiReact}) implementations of \textsc{3DReact} with the learning curves of the two models
presented in Figure~\ref{fig:learning_curves}.
Previous studies\cite{batzner_e3nn_2022, mace_2022} demonstrated that the equivariant models
showed superior extrapolation capabilities on predictions of energies and forces, as well as steeper and shifted learning
curves in force prediction tasks. Instead, we find that \textsc{InReact} and \textsc{EquiReact} are practically
indistinguishable for the present chemical reaction tasks.
}

\begin{table}[t!]
\begin{tabular}{@{}cccc@{}} \toprule
\makecell{Dataset \\ (property, units)} & \makecell{Atom-mapping\\ regime}
            & \textsc{InReact} & \textsc{EquiReact} \\
\midrule
\multicolumn{4}{@{}c@{}}{\emph{Random splits}}\\ \midrule
\multirow{3}{*}{\makecell{\gdb \\ ($\Delta E^\ddag$, kcal/mol)}}
& True & $ 4.93 \pm 0.18 $ & $ 4.93 \pm 0.15 $ \\
& RXNMapper & $ 6.03 \pm 0.26 $ & $ 6.05 \pm 0.25 $ \\
& None & $ 6.56 \pm 0.26 $ & $ 6.53 \pm 0.28 $ \\
\\[0.002cm]\multirow{3}{*}{\makecell{\cyclo \\ ($\Delta G^\ddag$, kcal/mol)}}
& True & $ 2.39 \pm 0.08 $ & $ 2.30 \pm 0.09 $ \\
& RXNMapper & $ 2.37 \pm 0.07 $ & $ 2.35 \pm 0.12 $ \\
& None & $ 2.39 \pm 0.05 $ & $ 2.31 \pm 0.09 $ \\
\\[0.002cm] \multirow{2}{*}{\makecell{\proparg \\ ($\Delta E^\ddag$, kcal/mol)}}
& True & $ 0.33 \pm 0.07 $ & $ 0.31 \pm 0.05 $ \\
& None & $ 0.34 \pm 0.06 $ & $ 0.31 \pm 0.06 $ \\
\midrule
\multicolumn{4}{@{}c@{}}{\emph{Scaffold splits}}\\ \midrule
\multirow{3}{*}{\makecell{\gdb \\ ($\Delta E^\ddag$, kcal/mol)}}
& True & $ 7.8 \pm 0.7 $ & $ 7.8 \pm 0.8 $ \\
& RXNMapper & $ 9.2 \pm 0.8 $ & $ 9.1 \pm 0.8 $ \\
& None & $ 10.1 \pm 0.9 $ & $ 10.0 \pm 0.9 $ \\
\\[0.002cm]\multirow{3}{*}{\makecell{\cyclo \\ ($\Delta G^\ddag$, kcal/mol)}}
& True & $ 2.79 \pm 0.18 $ & $ 2.72 \pm 0.18 $ \\
& RXNMapper & $ 2.77 \pm 0.22 $ & $ 2.71 \pm 0.23 $ \\
& None & $ 2.76 \pm 0.22 $ & $ 2.72 \pm 0.19 $ \\
\\[0.002cm] \multirow{2}{*}{\makecell{\proparg \\ ($\Delta E^\ddag$, kcal/mol)}}
& True & $ 0.44 \pm 0.11 $ & $ 0.40 \pm 0.08 $ \\
& None & $ 0.45 \pm 0.10 $ & $ 0.41 \pm 0.09 $ \\
\bottomrule
\end{tabular}
\caption{
Performance as measured in mean absolute errors (MAEs) of predictions
of \textsc{3DReact} (\textsc{InReact} \vs \textsc{EquiReact}).
\textsc{3DReact}$_M$ is used for the ``True'' and ``RXNMapper'' regimes,
and \textsc{3DReact}$_S$ is used for the ``None'' regime.
MAEs are averaged over 10~folds of 80/10/10 splits (training/validation/test)
and reported together with standard deviations across folds.
}
\label{tab:model_performance_equireact}
\end{table}

\begin{figure}[p]
\centering
\includegraphics[width=0.9\linewidth]{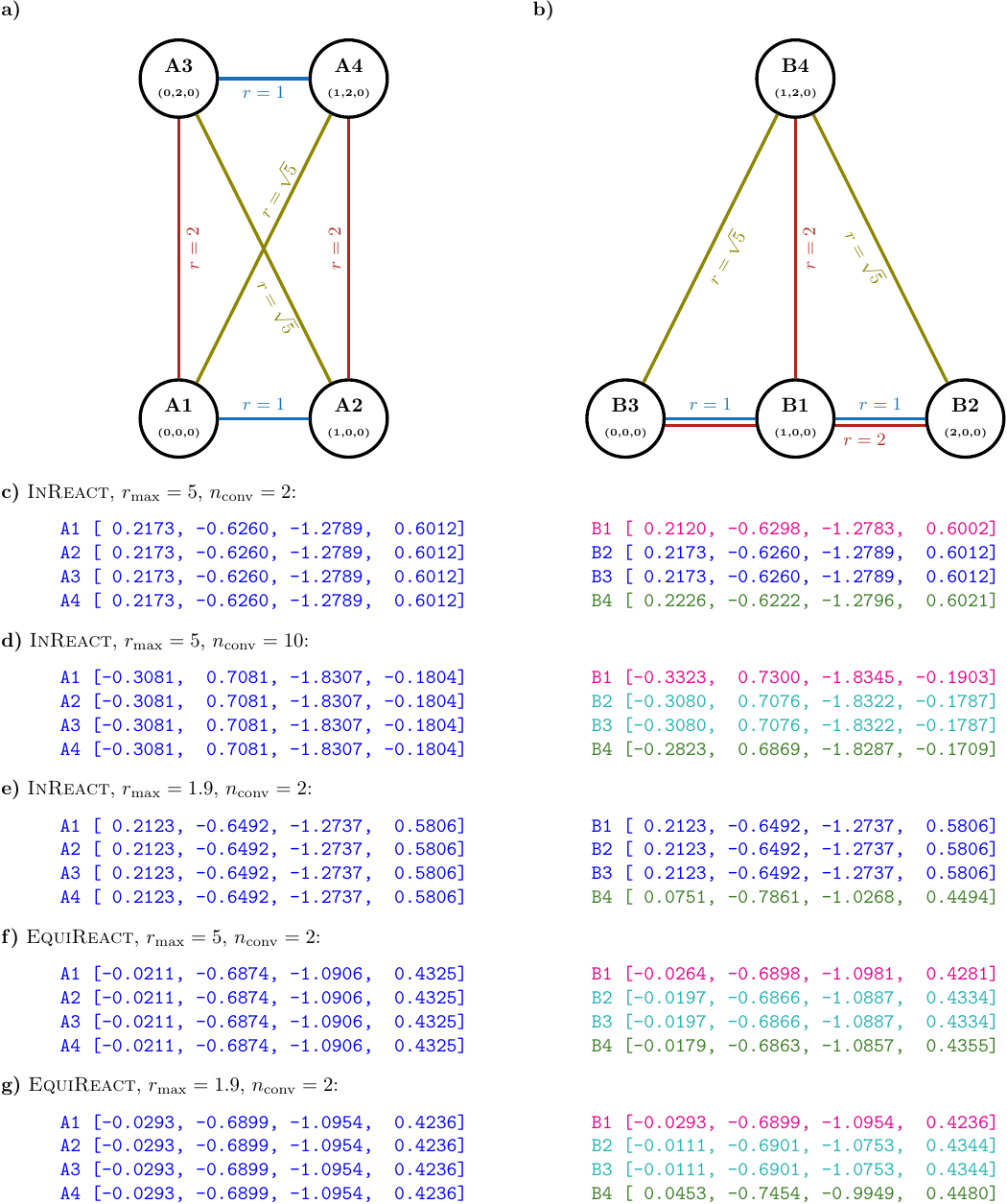}
\caption{
Top: Reactant and product of a toy reaction: two homometric structures (a) and (b) with atom labels,
atom coordinates (\AA), and interatomic distances (\AA).
``Bonds'' of the same length are of the same color.
Bottom: output after 5~epochs of the invariant (c,d,e) / equivariant (f,g) molecular channels for each atom
with different radial cutoffs $r_{\max}$
and number of convolutional layers $n_\mathrm{conv}$.
%($n_s=4$, $n_v=48$, $n_g=16$, $n_{\mathrm{neigh}}=20$, $p_d=0$).
Within each subfigure, atomic representations indistinguishable up to shown digits are marked by the same color.
}
\label{fig:homometric}
\end{figure}
\changed{We find that the datasets studied herein do not benefit from the inclusion of equivariant features for molecules.
Yet, Figure~\ref{fig:homometric} illustrates that a hypothetical reaction involving conversion
between homometric structures of \ce{He4},\cite{von_Lilienfeld_2013} which is mostly characterized by angle changes,
clearly benefits from equivariant molecular features.
In the reactant (Figure~\ref{fig:homometric}a), all atoms are identical and lead to the same learned representation.
In the product (Figure~\ref{fig:homometric}b), only atoms B2 and B3 have identical environments, different from A1--4.
Atoms B2--3 have the same distances $r$ to the three neighbors, as in A1--4.
Thus, \textsc{InReact}, which uses only interatomic distances,
yields very close representations for these
atoms (Figure~\ref{fig:homometric}c).
Still, in each convolutional layer, atoms B2 and B3 receive
information from B1 and B4,
and with increase of $n_\mathrm{conv}$ the difference in the representations
of B2--3 and A1--A4 becomes more apparent (Figure~\ref{fig:homometric}d).
However, with smaller radial cutoff $r_{\max}=1.9$,
atoms B1--B3 and A1--4 become indistinguishable for any number of layers (Figure~\ref{fig:homometric}e).
On the other hand, \textsc{EquiReact}, which uses explicit angular information from the spherical harmonics filters,
clearly distinguishes all non-equivalent atoms in both cases
already for $n_\mathrm{conv}=2$ (Figure~\ref{fig:homometric}f,g).}

\changed{While this is a toy example, it illustrates that transformations consisting of changes in angles rather than in bond
lengths are better described using \textsc{EquiReact}. In general, the currently available reaction datasets do not pose
sufficient challenge to allow distinguishing \textsc{InReact} and \textsc{EquiReact}. For the datasets studied in this work,
\textsc{InReact} is sufficient and is the model variation used throughout as \textsc{3DReact}.}

\clearpage
%%%%%%%%%%%%%%%%%%%%%%%%%%%%%%%%%%%%%%%%%%%%%%%%%%%%%%%%%%%%%%%%%%%%%%%%%%%%%%%%%%%%%%%%
\subsection{Benchmark studies}
\textsc{3DReact} is compared to previously best baseline models:\cite{digdisc}
\textsc{ChemProp},\cite{Heid2021, heid2023chemprop}
a graph neural network that uses atom-mapped SMILES to construct a CGR,
and the 3D-structure-based SLATM\cite{amons2020}
fingerprint adapted to reactions by taking the difference
between product and reactant fingerprints (SLATM$_d$),\cite{vangerwen_2022}
combined with KRR models (SLATM$_d$+KRR).

\changed{Note that both \textsc{3DReact} and \textsc{ChemProp} are run without explicit H atoms, for two reasons. First,
hydrogen atoms are not always mapped in the ``True'' and ``RXNMapper'' regimes, since they are usually implicit in SMILES
strings. Second, there is no consistent improvement in including H atoms in the models
(Table~\extref{tab:models-with-withoutH}). SLATM$_d$, built directly from molecular coordinates without using SMILES strings,
does however incorporate H atoms by default. For further discussion refer to Section~\extref{sec:hydrogens}.}

\subsubsection{Random splits}
\label{sec:results:performance}
Performance as measured in mean absolute errors (MAEs) is
illustrated in Table~\ref{tab:model_performance_all} for random splits of each dataset,
demonstrating the models' interpolative capabilities.
\changed{For the equivalent results with root mean squared errors (RMSEs), consult Section~\extref{sec:rmse}.}

\begin{table}[th!]
\begin{tabular}{@{}ccccc@{}} \toprule
\makecell{Dataset \\ (property, units)} & \makecell{Atom-mapping\\ regime}
            & \textsc{ChemProp} & SLATM$_d$+KRR & \textsc{3DReact} \\
\midrule
\multirow{3}{*}{\makecell{\gdb \\ ($\Delta E^\ddag$, kcal/mol)}}
& True & $\bf 4.35 \pm 0.15 $ & --- & $ 4.93 \pm 0.18 $ \\
& RXNMapper & $\bf  5.69 \pm 0.17 $ & --- & $\bf 6.03 \pm 0.26 $ \\
& None & $ 9.04 \pm 0.21 $ & $\bf 6.89 \pm 0.20 $ & $\bf 6.56 \pm 0.26 $ \\
\\[0.002cm]\multirow{3}{*}{\makecell{\cyclo \\ ($\Delta G^\ddag$, kcal/mol)}}
& True & $ 2.69 \pm 0.10 $ & --- & $\bf 2.39 \pm 0.08 $ \\
& RXNMapper & $ 2.71 \pm 0.07 $ & --- & $\bf 2.37 \pm 0.07 $ \\
& None & $ 2.71 \pm 0.12 $ & $ 2.65 \pm 0.08 $ & $ \bf 2.39 \pm 0.05 $ \\
\\[0.002cm] \multirow{2}{*}{\makecell{\proparg \\ ($\Delta E^\ddag$, kcal/mol)}}
& True & $ 1.53 \pm 0.14 $ & --- & $\bf 0.33 \pm 0.07 $ \\
& None & $ 1.56 \pm 0.16 $ & $\bf 0.33 \pm 0.04 $ & $\bf 0.34 \pm 0.06 $ \\
\bottomrule
\end{tabular}
\caption{
Performance as measured in mean absolute errors (MAEs) of predictions
of \textsc{3DReact} \vs state-of-the-art baselines \textsc{ChemProp} and SLATM$_d$+KRR.
All datasets are compared in three atom-mapping regimes:
``True'', ``RXNMapper'' and ``None'',
except for the \proparg set, where \textsc{RXNMapper} cannot map the reaction SMILES.
MAEs are averaged over 10~folds of random \changed{80/10/10} splits (training/validation/test)
and reported together with standard deviations across folds.
The lowest errors for each regime and dataset are highlighted in bold, if statistically relevant.
}
\label{tab:model_performance_all}
\end{table}

%%%%%%%%%%%%%%%%%%%%%%%%%%%%%%%%%%%%%%%%%%%%%%%%%%%%%%%%%%%%%%%%%%%%%%%%%%%%%%%%%%%%%%%%%%%%%%%%%%%%
The \gdb dataset is distinct from the other two
in that it includes variations in the reaction class (and mechanism),
thereby showing a greater dependence on the existence
and quality of atom-mapping information in the models.
It has already been observed\cite{digdisc}
for \textsc{ChemProp} that there is stark hierarchy in the predictions
from the ``True'' to ``RXNMapper'' to ``None'' regimes.

\begin{sloppypar}
In the ``True'' regime, \textsc{3DReact} does not improve predictive
capabilities over the \textsc{ChemProp} model for the \gdb set. This points to the
importance of the chemical diversity in this dataset, where knowledge of the
reaction mechanism (in the form of atom-maps) is sufficient information to
predict the reaction barriers without information about the
geometries of reactants and products.
\changed{However, as previously discussed,\cite{digdisc} ``True'' maps are an unrealistic scenario for most datasets.}
Moving to the ``RXNMapper'' regime, \textsc{3DReact} and \textsc{ChemProp} already
agree within standard deviations.
\changed{This highlights that for practical-quality maps, \textsc{3DReact} is amongst the best models for this dataset.}
In the ``None''
regime, \textsc{3DReact} outperforms \textsc{ChemProp} by more than
\SI{2}{kcal/mol}.
\end{sloppypar}

SLATM$_d$+KRR results in similar performance to \textsc{3DReact} for the \gdb set. The
SLATM$_d$ representation also constructs features from 3D coordinates of the reactants and
products using invariant functions, and is therefore more fundamentally similar to \textsc{3DReact} than \textsc{ChemProp}.
\changed{Nevertheless, since \textsc{3DReact} allows for the inclusion of atom-mapping information,
predictions are improved in the mapped regimes compared to SLATM$_d$+KRR, which operates in the ``None'' regime only.}

\changed{In summary, for the chemically diverse \gdb set,
while SLATM$_d$ allows for good performance in the ``None'' regime, and \textsc{ChemProp} in the ``True'' and ``RXNMapper'' regimes,
since \textsc{3DReact} can incorporate both atom-mapping information and 3D structure information,
the model achieves robust performance in all three regimes, with the predicted MAEs ranging from \SIrange{4.93}{6.56}{kcal/mol}}.

%%%%%%%%%%%%%%%%%%%%%%%%%%%%%%%%%%%%%%%%%%%%%%%%%%%%%%%%%%%%%%%%%%%%%%%%%%%%%%%%%%%%%%%%%%%%%%%%%%%%
The \cyclo\cite{coley_dataset} dataset contains
a single reaction class and has been previously illustrated\cite{digdisc}
to show less dependence on the quality of atom-mapping than the \gdb.
For this set, \textsc{3DReact} outperforms or matches the other models in all
three regimes.
This illustrates that a
model based purely on geometry information of reactants and products, without
any chemical information in the form of atom-mapping or surrogates thereof, can allow for accurate reaction property
prediction. \changed{It is worth noting that atom-mapping does not improve predictions at all, \ie there is no improvement from
``None'' to ``RXNMapper'' to ``True'', even for the \textsc{ChemProp} model. This points to the different nature of this
dataset compared to the \gdb}.

The best model is obtained with \textsc{3DReact}$_S$
in the \texttt{energy} mode (Figure~\ref{fig:combine}d). As outlined in
Section~\ref{sec:arch:reaction}, in \texttt{energy} mode an energy contribution
is learned for reactants' and products' atoms separately. In the original
publication,\cite{coley_dataset} Stuyver \etal illustrate that the
activation barriers ($\Delta G^\ddag$) correlate linearly with the reaction energy ($\Delta G$).
Since $\Delta G$ is the difference between products' and reactants' energies,
the \texttt{energy} mode is the best choice for a model learning the reaction energy,
and in the case of this dataset, for $\Delta G^\ddag$ too, due to its linear correlation with $\Delta G$.

\changed{Compared to SLATM$_d$+KRR, \textsc{3DReact} in the ``None'' regime results in lower prediction errors for this set,
illustrating that despite both models using similar information, an end-to-end model can allow for improved predictions.}

%%%%%%%%%%%%%%%%%%%%%%%%%%%%%%%%%%%%%%%%%%%%%%%%%%%%%%%%%%%%%%%%%%%%%%%%%%%%%%%%%%%%%%%%%%%%%%%%%%%%
The \proparg\cite{Doney2016, Gallarati2021} is a small dataset for neural network standards
(753~points) and therefore constitutes a challenge for the data efficiency of our model.
Like the \cyclo set, it consists of a single reaction class,
\ie enantioselective propargylation of benzaldehyde.
Since the enantioselectivity is related to the barrier through an exponential relationship,
it is critical to predict the barrier accurately ($\leq$~\SI{1}{kcal/mol}).\cite{Gallarati2021}
The ``RXNMapper'' regime is not available since \textsc{RXNMapper} cannot atom-map the reaction SMILES of this set.

In the other regimes, 3D-structure-based models lead to the best results, outperforming \textsc{ChemProp} by a large margin.
\proparg is particularly hard for 2D-based models\cite{digdisc}
since it contains molecules of different stereochemistry but the same SMILES strings.
Again trained on a single-reaction class dataset, models do not benefit from being provided the ``obvious'' chemical information:
including true atom-maps does not decrease the error.
\changed{Competing only in the ``None'' regime,
\textsc{3DReact} does not allow for a performance improvement compared to SLATM$_d$+KRR.
Given the small size of the dataset, it is already a demonstration of data efficiency
that the deep-learning model matches the prediction errors of the kernel model.
Unlike for \textsc{Nequip}\cite{batzner_e3nn_2022} however,
the data efficiency here is not due to the equivariant molecular components (Section~\ref{sec:results:eq_vs_inv}).}

\bigskip

The three datasets illustrate the benefits of the flexibility of \textsc{3DReact}:
depending on the datasets' particular challenges,
the model exploits the available information to yield the best-performing model
in almost all cases. Since the model settings (such as \texttt{vector} or \texttt{energy} mode choice)
are specified as hyperparameters,
the optimized version of \textsc{3DReact} can emerge with minimal user intervention.

%%%%%%%%%%%%%%%%%%%%%%%%%%%%%%%%%%%%%%%%%%%%%%%%%%%%%%%%%%%%%%%%%%%%%%%%%%%%%%%%%%%%%%%%%%%%%%%%%%%%
\subsubsection{\changed{Extrapolative splits}}
\label{sec:results:extrapolation}
\begin{figure}[tbh!]
\centering
\includegraphics[width=\textwidth]{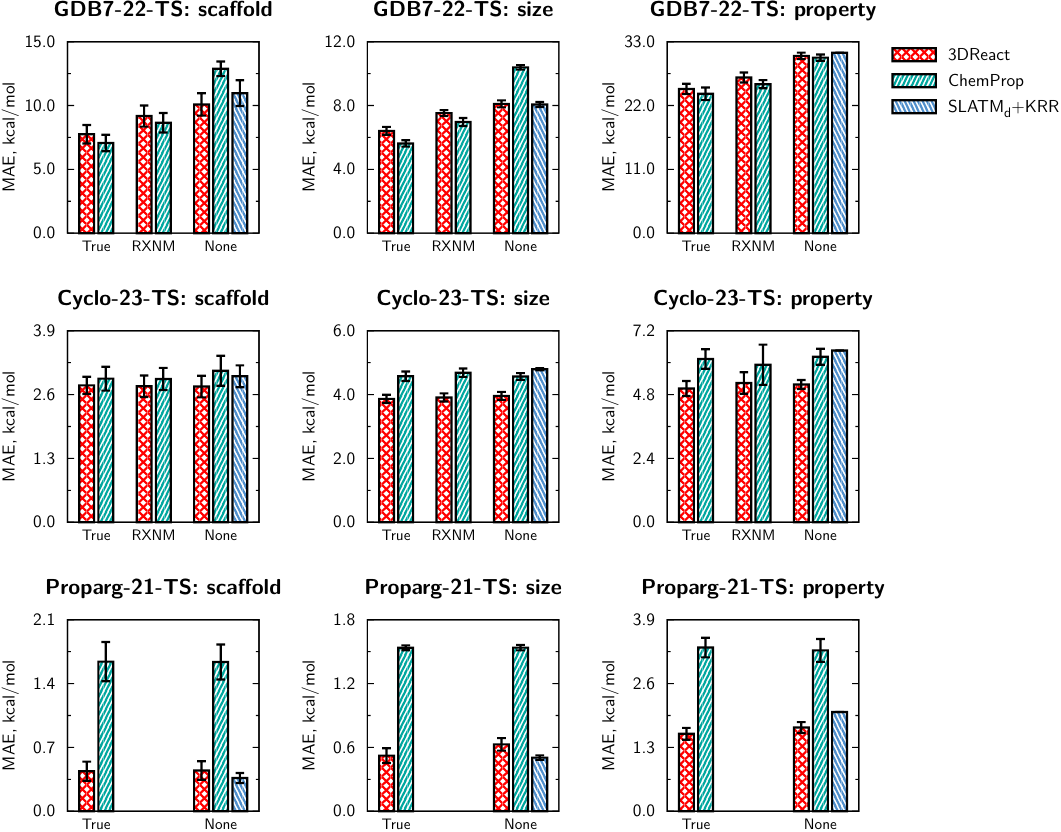}
\caption{\changed{Mean absolute errors (MAEs) of predictions
using three different extrapolation splits: scaffold, size-, and property-based.
All datasets are compared in three atom-mapping regimes:
``True'', ``RXNMapper'' (RXNM), and ``None'',
except for the \proparg set, where \textsc{RXNMapper} cannot map the reaction SMILES.
MAEs are averaged over 10~folds of 80/10/10 splits (training/validation/test),
and error bars indicate standard deviations across folds, where applicable.}
}
\label{fig:extrapolation_splits}
\end{figure}

\changed{Figure~\ref{fig:extrapolation_splits} illustrates model performance for extrapolative splits (based on scaffolds,
molecular size of reactants/products, and barrier magnitude, detailed in Section~\ref{sec:methods:splits}).
These different types of extrapolative splits are necessarily more difficult than random splits, as demonstrated by higher MAEs
in Figure~\ref{fig:extrapolation_splits}. The relative performance of the models is largely maintained in the three different
extrapolation regimes compared to the interpolation regime presented in Table~\ref{tab:model_performance_all}.}

Bemis--Murcko scaffold\cite{Bemis_1996} splitting clusters molecules
(reactants for \gdb and \proparg, products for \cyclo) based on ring systems.
Test molecules may therefore appear ``novel'' from the point of view of the reaction graph,
but will still feature distances and angles close to what the model has seen during training.
\changed{Similarly for size-based splits, since there is no correlation between reactant/product size and reaction barriers,
using distance information allows for stable predictions on extrapolation.
Property-based splits are more challenging than the other two. For the \cyclo and \proparg sets, \textsc{3DReact} still offers
respectable errors, lower than those of the other models. For the \gdb set however, all models result in unreasonable MAEs over
\SI{20}{kcal/mol}. This points to the particular challenges of the \gdb set and suggests an avenue for further developments of
ML models for extrapolative tasks.\cite{Vadaddi_2024}}

\changed{Again in contrast to previous works that suggested equivariant models might be better at extrapolation
tasks,\cite{batzner_e3nn_2022, mace_2022} here we find that \textsc{3DReact} offers stable extrapolation performance
(particularly for size- and scaffold-based splits), but not necessarily improved extrapolation behavior compared to 2D-graph
based models. This points to the different challenges in reaction property prediction. Nevertheless,
Figure~\ref{fig:extrapolation_splits} illustrates that \textsc{3DReact} is a consistently robust model for the three datasets
when moving from interpolation to extrapolation regimes.}

%%%%%%%%%%%%%%%%%%%%%%%%%%%%%%%%%%%%%%%%%%%%%%%%%%%%%%%%%%%%%%%%%%%%%%%%%%%%%%%%%%%%%%%%%%%%%%%%%%%%
\subsection{Model behavior}
\label{sec:results:model_behaviour}
Since the \gdb set has the largest chemical diversity amongst the datasets explored, studying \textsc{3DReact}
and baseline models SLATM$_d$ and \textsc{ChemProp} on this dataset
best captures the different chemical interpretation provided by these models.

\begin{figure}[tb]
\centering
\includegraphics[width=\linewidth]{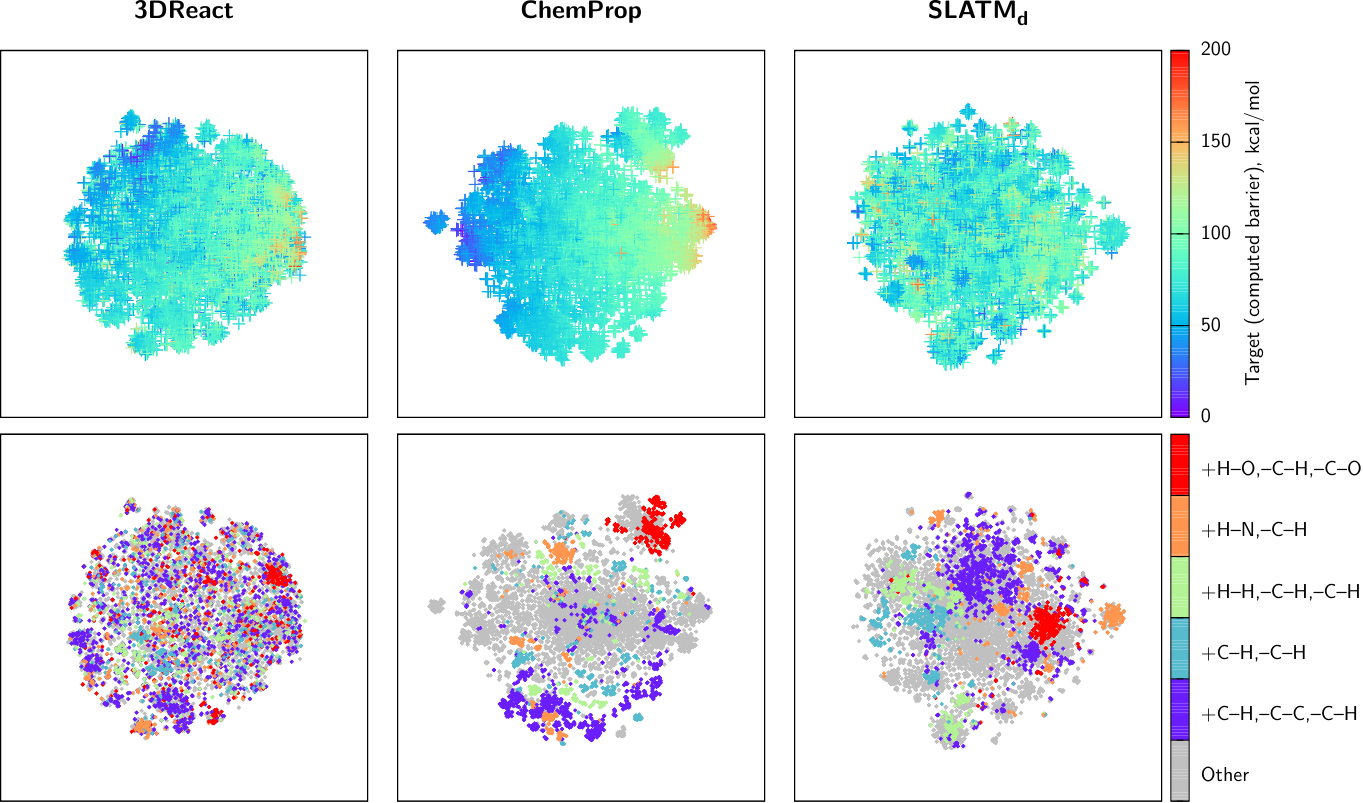}
\caption{
t-SNE maps (perplexity $=64$)
of the latent representations of \textsc{3DReact} and \textsc{ChemProp} models in the ``True'' regime
and the SLATM$_d$ representation of the \gdb dataset,
colored by the target $\Delta E^\ddag$ (upper panels) and reaction types (lower panels).
}
\label{fig:tsnes}
\end{figure}

Figure~\ref{fig:tsnes} compares
the (latent) representations
of \textsc{3DReact} ``True'', \textsc{ChemProp} ``True'' and SLATM$_d$
using t-SNE\cite{van2008visualizing} maps.
In the upper panel, we find that the quality of the correlation between
the representations and the target property
are aligned with the relative performance of the models
(Table~\ref{tab:model_performance_all}).
\textsc{ChemProp} and \textsc{3DReact} show a smooth transition
of the target property,
whereas the map of SLATM$_d$ does not have a clear structure.
The lower panel shows the correlation of the representations
with the five most common reaction types
defined by bond breaking and formation (see Section~\ref{sec:methods:xtb}).
\textsc{ChemProp}, as a chemically-inspired model,
illustrates clear clusters in the reaction type.
While SLATM$_d$ is a geometry-based model,
the binning structure used to create the representation\cite{amons2020, vangerwen_2022}
results in a clear correlation with the reaction types,
since \eg the pairwise bins
naturally cluster features such as \ce{C}--\ce{H}
bond formation or breaking.
\changed{\textsc{3DReact} shows the least distinct ``chemical'' clustering,
due to the interplay of geometry and mapping information exploited in the representation.}

\begin{figure}[tb]
\centering
\includegraphics[width=0.62\linewidth]{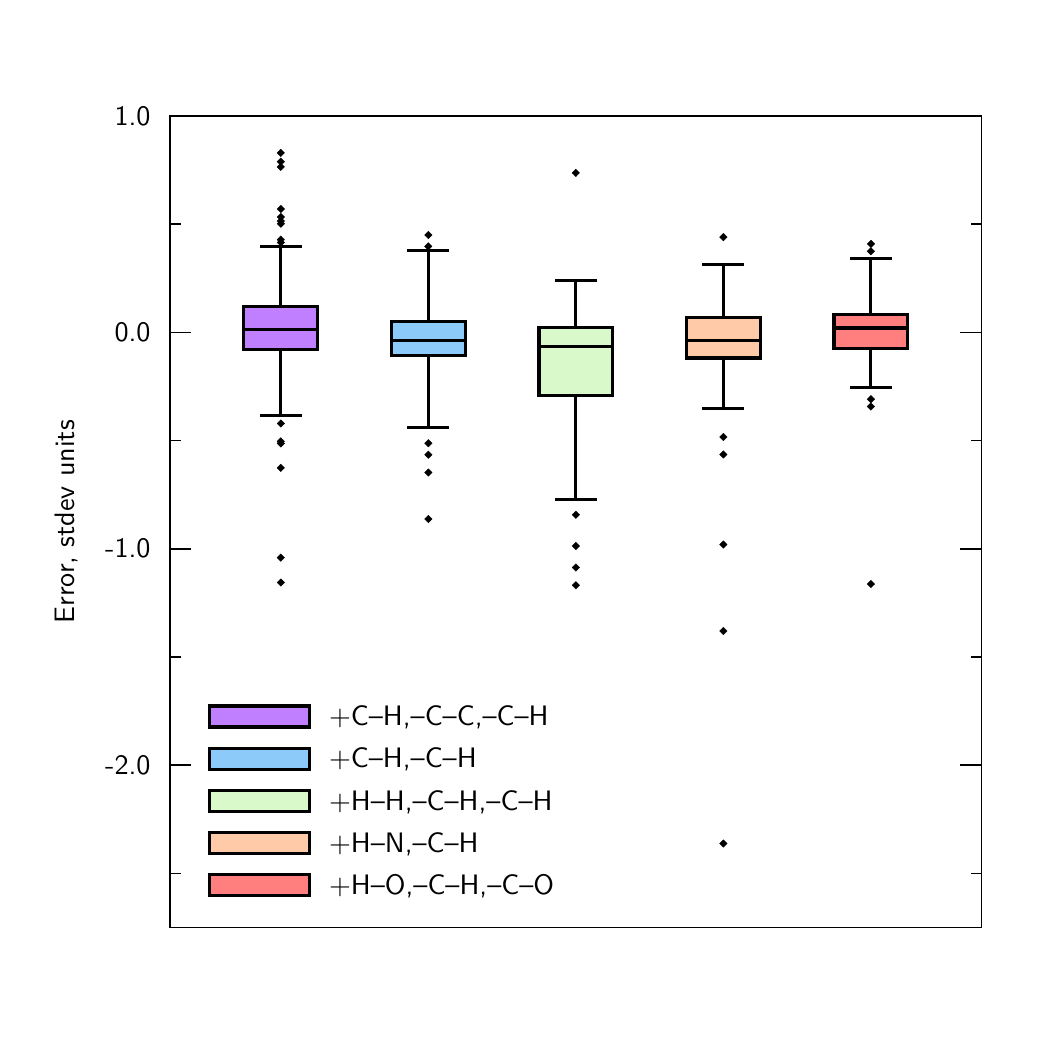}
\caption{
Box plots illustrating how \textsc{3DReact} ``True'' performs
for the most common reaction types in the \gdb set.
\textsc{3DReact} is constructed without explicit \ce{H} nodes in the graphs.
The boxes range from the first to the third quartile of the datapoints.
The whiskers limit 90\% of the datapoints and the individual points illustrate outliers.
The points correspond to the test set of the first random split.
The errors are given in the target standard deviation (stdev) units (\SI{21.8}{kcal/mol}).
}
\label{fig:box_plot_reaction_classes}
\end{figure}

Figure~\ref{fig:box_plot_reaction_classes}
shows the error distribution of predictions belonging to the same reaction classes
for \textsc{3DReact} ``True''.
\textsc{3DReact} performs universally well across the different reaction types,
with consistently low errors and relatively small error spread.
The reactions for which the model has higher mean errors and spread
(\mbox{$+$H--H,$-$C--H,$-$C--H} (green)) correspond to those involving C--H and H--H features.
Since the model is trained without explicit H nodes in the graph,
features associated with X--H bonds are included implicitly in the model.
\changed{Capturing H--H bond changes will be the most challenging as these will be the least explicitly described,
occurring only as initial features for neighboring nodes.}
Since C is the most frequently occurring element in various different configurations,
capturing all the C--H features is more challenging than the O--H features for example, which will be more similar to one another.
The equivalent plot for the model trained with explicit H nodes
is shown in Figure~\extref{fig:box_plot_with_H},
illustrating that the error spread reduces
for the reaction types involving C--H and H--H features.
Note that \textsc{3DReact} without explicit Hs still leads to performance
comparable to the variant with explicit Hs (Section~\extref{sec:hydrogens}).

%%%%%%%%%%%%%%%%%%%%%%%%%%%%%%%%%%%%%%%%%%%%%%%%%%%%%%%%%%%%%%%%%%%%%%%%%%%%%%%%%%%%%%%%%%%%%%%%%%%%
\subsection{Geometry quality}
\label{sec:results:geometry-quality}
\begin{figure}[tb]
\centering
\includegraphics[width=0.95\textwidth]{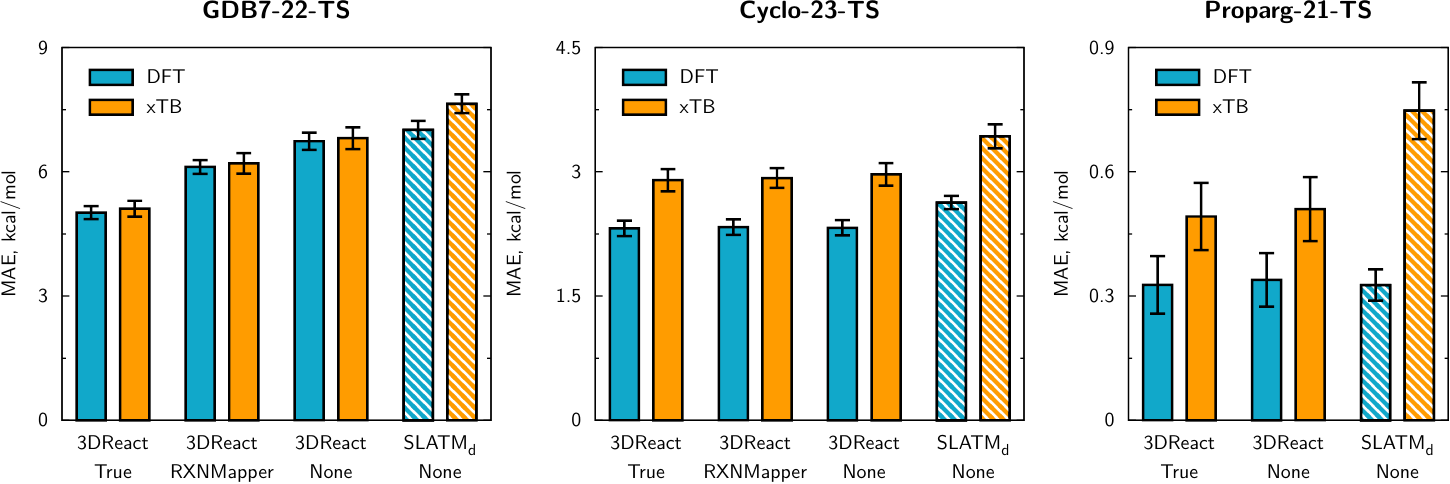}
\caption{
Mean absolute errors (MAEs) for predictions
using either the provided geometries
($\omega$B97X-D3/def2-TZVP for \gdb,
B3LYP-D3(BJ)/def2-SVP for \cyclo,
B97D/TZV(2p,2d) for \proparg) (DFT)
or lower-quality GFN2-xTB (xTB)
geometries.
MAEs are averaged over 10 folds of random \changed{80/10/10} splits (training/validation/test),
error bars showing standard deviations across folds.
Note that for \gdb and \cyclo datasets
the DFT results are different from those presented in Section~\ref{sec:results:performance}
because here they are obtained on the same subset as the xTB results
(see Section~\ref{sec:methods:xtb}).}
\label{fig:xtb_errors}
\end{figure}

\begin{sloppypar}
In order to illustrate that \textsc{3DReact} does not require high-quality molecular structures
to be used in an out-of-sample scenario,
we train and test a model using lower-quality GFN2-xTB\cite{gfn2_xtb} (xTB)
geometries to predict higher-level barriers
(CCSD(T)-F12a/cc-pVDZ-F12//$\omega$B97X-D3/def2-TZVP for \gdb,
B3LYP-D3(BJ)/def2-TZVP//B3LYP-D3(BJ)/def2-SVP for \cyclo
and B97D/TZV(2p,2d) for \proparg).
The results are illustrated in Figure~\ref{fig:xtb_errors}
for the three datasets with DFT and xTB geometries,
and compared to the SLATM$_d$+KRR model in the same settings.
\textsc{3DReact} benefits from a lower sensitivity to the geometry quality
compared to the pre-designed representation SLATM$_d$ combined with KRR, across the three datasets.
\end{sloppypar}

For the \gdb set, there is a
negligible difference in model performance moving from DFT to xTB geometries.
The xTB geometries are a good proxy for the DFT
ones here, since this set consists of small, charge-neutral organic
molecules, which are largely well-described by semi-empirical methods.
For the \cyclo set, while the molecules are still organic, they are larger than
those in the \gdb set, and there is a greater divergence between the GFN2-xTB
and DFT geometries, resulting in a larger deterioration with these structures.
Figure~\extref{fig:cyclo_rmsd} demonstrates that when using the model trained on xTB geometries,
barrier predictions for molecules with poorer geometries (\ie, higher RMSD of xTB \vs DFT geometries)
are not necessarily worse than those on molecules with better geometries.
Instead, there is a consistent decline in model performance when training with xTB geometries and predicting DFT barriers.

The \proparg set is the most complex of the three for GFN2-xTB, since these
systems with charged organosilicon compounds differ considerably from those
used to parameterize semi-empirical methods or force fields.
As described in Section~\ref{sec:methods:xtb}, unlike for the other datasets
where we generate an initial structure from SMILES using force fields,
for this set it is impossible and we instead generate xTB geometries from the DFT ones.
While this is not a feasible geometry generation pipeline for out-of-sample predictions,
it still demonstrates how different methods perform with high and low-quality geometries.
Here, we see that \textsc{3DReact} is less sensitive than SLATM$_d$+KRR
and the variation trained with lower quality geometries still offers competitive errors
($0.48 \pm 0.05$~kcal/mol for the ``None'' model).

%%%%%%%%%%%%%%%%%%%%%%%%%%%%%%%%%%%%%%%%%%%%%%%%%%%%%%%%%%%%%%%%%%%%%%%%%%%%%%%%%%%%%%%%%%%%%%%%%%%%
\section{Conclusions}

\changed{The accurate and reliable prediction of reaction barriers across diverse sets of chemical reactions remains an open
challenge in computational chemistry.}
We contribute to
this domain by introducing \textsc{3DReact}, a \changed{geometric deep learning model} constructed
from the 3D coordinates of reactants and products.
\changed{We show that the invariant model (\vs the equivariant version) is already sufficient for currently available reaction datasets.
Existing models \textsc{ChemProp} and SLATM$_d$+KRR exhibit impressive performance for atom-mapped, chemically diverse datasets
and stereochemistry-sensitive datasets, respectively. \textsc{3DReact} offers a hybrid model that can optionally incorporate
mapping information alongside geometries, enabling robust performance across different dataset types and atom-mapping regimes.
\textsc{3DReact} also allows for a reduced sensitivity to the training geometry quality (\ie, xTB \vs DFT level) compared to
SLATM$_d$+KRR.
Predictions are stable both when moving to molecular size- or scaffold-based splits.
Altogether, \textsc{3DReact} presents a flexible framework for accurate prediction of activation barriers across chemical
reaction datasets.}
\changedagain{Despite the proposed developments, challenges remain for ML predictions of energy barriers, particularly in
integrating them within experimental settings. This work is a step toward their reliable application.}

%%%%%%%%%%%%%%%%%%%%%%%%%%%%%%%%%%%%%%%%%%%%%%%%%%%%%%%%%%%%%%%%%%%%%%%%%%%%%%%%%%%%%%%%%%%%%%%%%%%%
\section{Methods}

\subsection{Datasets}
\label{sec:methods:datasets}

We test \textsc{3DReact} on three datasets of reaction barriers
previously used to benchmark reaction representations.\cite{digdisc}
\changed{The term ``reaction barrier'', used interchangeably with ``activation energy'' and ``activation barrier''
is the energy difference between the energy of the optimized TS and the optimized reactants. Note that depending on the dataset,
some provide purely electronic energies (labelled $\Delta E^\ddag$)
and others --- Gibbs free energies (labelled $\Delta G^\ddag$).}
In all datasets, optimized three-dimensional structures of reactants
and products are provided, which are used to train models and make
predictions. The activation barrier is not a direct function of these
structures, but using the TS structure to make predictions removes the utility
of the ML models \vs direct computation of the TS. Thus we use an implicit interpolation of
reactants' and products' structures as a proxy for the TS as in previous
works.\cite{vangerwen_2022, Gallarati2021, digdisc}

The \gdb\cite{gdb7-22-ts} dataset consists of
close to \num{12000} diverse organic reactions automatically constructed from the
GDB7 dataset\cite{gdb13, reymond2015chemical, ramakrishnan2014quantum} using
the growing string method\cite{zimmerman_growing_string} along with
corresponding energy barriers ($\Delta E^\ddag$) computed at the
CCSD(T)-F12a/cc-pVDZ-F12//$\omega$B97X-D3/def2-TZVP level.
The dataset provides atom-mapped SMILES, with ``True'' maps derived from the transition state.
For \num{43}~reactions out of \num{11926},
one of the products' SMILES represents a molecule
different from the xyz structure.
These reactions were therefore excluded from the dataset,
leading to a modified \gdb set used here.

While there are no pre-defined classes for all the reactions
in the GDB7-20-TS\cite{Grambow2020} or \gdb\cite{gdb7-22-ts} sets,
Grambow \etal\cite{grambow_deep_learning} split the dataset
into reactions undergoing certain bond changes:
for example,
the most common type
was breaking of a C--H bond ($-$C--H) and a C--C bond ($-$C--C) in the reactants
and formation of a C--H bond ($+$C--H) in the products, giving the
reaction type signature \mbox{$+$C--H,$-$C--C,$-$C--H}.
Here, we extract similar reaction types by
comparing the connectivity matrices from atom-mapped reaction SMILES
of reactants and products (ignoring bond orders).
The most abundant reaction types in the dataset are
\mbox{$+$C--H,$-$C--C,$-$C--H}   (1667~reactions),
\mbox{$+$H--N,$-$C--H}           (633),
\mbox{$+$C--H,$-$C--H}           (619),
\mbox{$+$H--O,$-$C--H,$-$C--O}   (599) and
\mbox{$+$H--H,$-$C--H,$-$C--H}   (517).

The original \cyclo\cite{coley_dataset} dataset encompasses \num{5269}~profiles
for $[3+2]$~cycloaddition reactions with activation free energies ($\Delta G^\ddag$)
computed at the B3LYP-D3(BJ)/def2-TZVP//B3LYP-D3(BJ)/def2-SVP level
\changed{in water using the SMD continuum solvation model.}
The dataset provides atom-mapped SMILES with ``True'' maps
for heavy atoms derived from either the
transition state structure or heuristic rules.
For the regime with explicit hydrogen atoms,
we atom-mapped the xyz files by matching the reactants,
given in two separate files,
to the provided transition state structure,
which closely resembles the two reactants and
has the same atom order as in the product.
This was done with a labelled graph matching algorithm
as implemented in \texttt{NetworkX}.\cite{cordella2001improved,networkx}
The algorithm is unaware of chirality, double-bond stereochemistry
or conformations, and thus may lead to not exactly correct atom-mappings.
We also found that in four reactions,
the product SMILES and xyz files depict different species,
thus the set was reduced to \num{5265}~reactions.

The \proparg dataset\cite{Doney2016, Gallarati2021} contains 753~structures
of intermediates before and after the enantioselective transition state
of benzaldehyde propargylation,
with activation energies ($\Delta E^\ddag$) computed at the B97D/TZV(2p,2d) level.
SMILES strings (``fragment-based'' SMILES) and ``True'' atom-maps are not provided
with the original dataset, these are taken from Ref.~\onlinecite{digdisc}.

\begin{sloppypar}
\textsc{RXNMapper}\cite{schwaller2021extraction}-mapped versions of \gdb and \cyclo
were obtained with the python package \texttt{rxnmapper} (version 0.3.0), using the default settings.
The \proparg set cannot be mapped, because the underlying libraries cannot process its SMILES string.\cite{digdisc}
Since \textsc{RXNMapper} sorts molecules in case of multiple reactants and/or products,
which would complicate SMILES--xyz matching (see Section~\ref{sec:methods:smiles-matching} below),
we used a locally modified version that does not change the molecule order
(the patch file is provided in the project repository at
\url{https://github.com/lcmd-epfl/EquiReact/tree/9d78892fe/data-curation/rxnmapper}).
\end{sloppypar}

%%%%%%%%%%%%%%%%%%%%%%
\subsection{\changed{Data splits}}
\label{sec:methods:splits}

\changed{For each dataset and splitting type, identical data splits were used for all the models compared.}
\changed{In each case, ten different splits are constructed with different random seeds.}

\changed{Three different types of extrapolation split were used: scaffold-, molecular size- and property-based.
Scaffold splitting\cite{Wu_2018,yang2019analyzing} clusters molecules
based on their 2D backbones (such as Bemis--Murcko scaffolds\cite{Bemis_1996})
and ensures that the clusters (scaffolds) belonging to the training, validation, and test sets
do not overlap.
Size-based splitting organizes the splits such that the reactions of the smallest molecules are in the training set and the
reactions of the largest molecules are in validation and test. With property-based splits, one trains on reactions with higher
barriers and predicts on reactions with lower barriers. This choice of splits reflects the relevant out-of-sample cases: larger
molecules are more expensive to compute, and reactions with smaller barriers are desirable. Size- and property-based splits can
also be organized in reverse order, where larger molecules are in the train set and smaller in test, or reactions with lower
barrier in train and higher barrier in test.}

\changed{For molecular size- and scaffold-based splits, the initial data shuffling affects the composition of the datasets.
The non-zero standard deviations for property-based splits with neural networks
arise from different organization of the datapoints into batches.}

%%%%%%%%%%%%%%%%%%%%%%%%%%%%%%%%%%%%%%%%%%%%%%%%%%%%%%%%%%%%%%%%%%%%%%%%%%%%%%%%%%%%%%%%%%%%%%%%%%%
\subsection{Matching SMILES strings to xyz geometries}
\label{sec:methods:smiles-matching}

\textsc{3DReact} makes use of both the graph structure of a molecule (as provided in the SMILES string)
and the three-dimensional structure (in the xyz).
The atoms in the graph are associated with the atomic coordinates provided in the xyz file.
Thanks to the way the \gdb dataset\cite{gdb7-22-ts} was generated,
the atomic coordinates can be easily matched to SMILES
which in turn allows to atom-map reactants to products.
However, we also tested \textsc{RXNMapper}-mapped SMILES which do not respect the same
constraints. Therefore, for consistency, we use a SMILES--xyz matching
procedure detailed below.

We construct molecular graphs from xyz using covalent radii and matched them
to \texttt{RDKit}\cite{rdkit2023} molecular graphs obtained from SMILES
with a labelled graph matching algorithm as implemented in \texttt{NetworkX}.\cite{cordella2001improved,networkx}
This procedure is however unaware of chirality and double-bond stereochemistry,
thus some of the matches might be incorrect.
Still, it provides a flexible method that can be applied to any dataset consisting of SMILES strings and xyz files.

The same procedure was applied to the \cyclo dataset
in the few cases when the canonical SMILES
have a different atom ordering than xyz.

%%%%%%%%%%%%%%%%%%%%%%%%%%%%%%%%%%%%%%%%%%%%%%%%%%%%%%%%%%%%%%%%%%%%%%%%%%%%%%%%%%%%%%%%%%%%%%%%%%%%
\subsection{xTB geometry generation}
\label{sec:methods:xtb}

For the \gdb and \cyclo datasets,
the starting structures were generated from SMILES using the distance-geometry embedding
implemented in \texttt{RDKit}\cite{rdkit2023} with the srETKDGv3
settings.\cite{riniker2015better} Ten conformations were produced per molecule,
which were then energy-ranked with the MMFF94 implementation\cite{Tosco2014} in
\texttt{RDKit}, defaulting to UFF in case of missing parameters.
The lowest energy conformer was retained.
For the \proparg set, the original B97D/TZV(2p,2d) geometries
were used as a starting point, because the
stereochemical and conformational diversity of this set
cannot be completely encoded with SMILES. Therefore MMFF94 will fail to generate an initial geometry from SMILES.

For all the sets, the starting structures
were optimized at the GFN2-xTB semi-empirical level of
theory\cite{gfn2_xtb} at the ``loose'' convergence level for a maximum of
1000~iterations using \texttt{xTB}\cite{xtb_package} version 6.2 RC2.
For \num{969}~reactions of the \gdb set
and \num{491}~reactions of the \cyclo set,
at least one of the participating molecules either could not converge
to any reasonable configuration or converged to a structure not matching the SMILES.
These reactions were excluded from the geometry quality tests (Section~\ref{sec:results:geometry-quality}).

%%%%%%%%%%%%%%%%%%%%%%%%%%%%%%%%%%%%%%%%%%%%%%%%%%%%%%%%%%%%%%%%%%%%%%%%%%%%%%%%%%%%%%%%%%%%%%%%%%%
\subsection{Model training}
\textsc{3DReact} was trained using the Adam optimizer
\cite{kingma2014adam} with initial learning rate and weight decay parameters as hyperparameters.
The learning rate was reduced by 40\% after \num{60}~epochs
of no improvement in the validation MAE, as in Ref.~\onlinecite{stark2022equibind}.
Models were trained for max.\ \num{512}~epochs,
using early stopping after \num{150}~epochs of no improvement.
The model with the best validation score was then used to make predictions on the test set.

The optimal model hyperparameters were searched within the following values:
learning rate $\in [\num{5e-5}, \num{e-4}, \num{5e-4}, \num{e-3}]$;
weight decay parameter $\in [\num{e-5}, \num{e-4}, \num{e-3}, 0]$;
node and edge features embedding size $n_s \in [16, 32, 48, 64] $;
$\ell{=}1$ hidden space size $n_v \in [16, 32, 48, 64] $;
number of edge features $n_g \in [16, 32, 48, 64]$;
number of convolutional layers $n_\mathrm{conv} \in  [2, 3] $;
radial cutoff $r_{\max} \in [2.5, 5.0, 10.0] $;
maximum number of atom neighbors $n_{\mathrm{neigh}} \in [10, 25, 50]$;
dropout probability $p_d \in [0.0, 0.05, 0.1]$;
\texttt{sum\_mode} $\in$ [\texttt{node}, \texttt{both}];
\texttt{combine\_mode} $\in$ [\texttt{mlp}, \texttt{diff}, \texttt{mean}, \texttt{sum}];
\texttt{graph\_mode} $\in$ [\texttt{energy}, \texttt{vector}].

The hyperparameter search was done \changed{for the equivariant model} \textsc{EquiReact}$_S$
(without attention or mapping) using Bayesian search as implemented in Weights \& Biases.\cite{wandb}
Hydrogen atoms were \changed{excluded from} the graphs.
Sweeps were run for 128~epochs for the \gdb and \proparg sets,
and for 256~epochs for the \cyclo set on the first random split.
The parameters resulting in the best validation error, summarized in Table~\extref{tab:model-params},
were used for all the other model settings.

%%%%%%%%%%%%%%%%%%%%%%%%%%%%%%%%%%%%%%%%%%%%%%%%%%%%%%%%%%%%%%%%%%%%%%%%%%%%%%%%%%%%%%%%%%%%%%%%%%%%
\subsection{Baseline models}
The \textsc{ChemProp} model\cite{heid2023chemprop}
is based on a
CGR built from atom-mapped SMILES strings of reactants and products,
which is then passed through the directed message-passing neural network
\texttt{chemprop}\cite{yang2019analyzing,Heid2021,heid2023chemprop} (version 1.5.0).
The hyperparameters are taken from Ref.~\onlinecite{digdisc}.

Molecular SLATM vectors were generated using the \texttt{qml} python package\cite{qml}
before being combined to form the reaction version SLATM$_d$.
SLATM$_d$ is used with kernel ridge regression (KRR) models.
The kernel functions and widths, and regularization parameters, were optimized
on the first of the ten random splits,
in line with how the hyperparameters were optimized for \textsc{3DReact}. Unlike \textsc{3DReact},
the hyperparameters for DFT and xTB geometries
were optimized separately.

%%%%%%%%%%%%%%%%%%%%%%%%%%%%%%%%%%%%%%%%%%%%%%%%%%%%%%%%%%%%%%%%%%%%%%%%%%%%%%%%%%%%%%%%%%%%%%%%%%%%
\section*{Data and Software Availability statement}
The code is available as a GitHub repository at
\url{https://github.com/lcmd-epfl/EquiReact}.
The versions of the datasets used, as well as any processing applied to them, can be found in the same repository.
\changed{The unprocessed results are available in the same same repository
as well as at \url{https://wandb.ai/equireact}.}

\begin{suppinfo}
Supplementary Information is provided in the freely available file \texttt{equireact\_si.pdf},
detailing
\changed{the architecture of the molecular channels
(Section~\extref{sec:molecular_channels})},
the \changed{\textsc{3DReact}} hyperparameters
(Section~\extref{sec:model-params}),
\changed{the RMSE analogue of Table~\ref{tab:model_performance_all}
(Section~\extref{sec:rmse})},
the discussion of the model with a cross-attention surrogate for atom-mapping
(Section~\extref{sec:cross}),
\changed{extrapolation studies
(Section~\extref{sec:extrapolation})},
\changed{some illustrative correlation plots for the \gdb set
(Section~\extref{sec:gdb_outliers_corr})},
the model performance with and without explicit hydrogen atoms
(Section~\extref{sec:hydrogens}),
and the geometry sensitivity analysis for the \cyclo set
(Section~\extref{sec:geom_cyclo}).
\end{suppinfo}

\section*{Author Information}

\subsection*{Author contributions}
P.v.G., K.R.B., and C.B. conceptualized the project.
\textsc{3DReact} and support codes were written and run by K.R.B. and P.v.G.,
with design suggestions from C.B., V.R.S., and R.L.
Results were analyzed by P.v.G., K.R.B., V.R.S., R.L., and C.B.
xTB computations were run by R.L.
The original draft was written by P.v.G. and K.R.B. with reviews and edits from all authors.
C.C. and A.K. provided supervision and acquired funding.

\subsection*{Conflict of interest}
The authors have no conflicts to disclose.

\begin{acknowledgement}
The authors thank Liam Marsh and Yannick Calvino Alonso
for helpful discussion and comments on the text.
P.v.G., C.B., V.R.S., R.L., A.K., and C.C. acknowledge the National Centre of Competence in Research (NCCR)
``Sustainable chemical process through catalysis (Catalysis)'', grant number~180544,
of the Swiss National Science Foundation (SNSF) for financial support.
K.R.B. and C.C. were supported by the European Research Council (grant number~817977)
and by
the National Centre of Competence in Research (NCCR)
``Materials' Revolution: Computational Design and Discovery of Novel Materials (MARVEL)'',
grant number~205602,
of the Swiss National Science Foundation.
\end{acknowledgement}

\bibliography{equireact.bib}

\clearpage

\begin{tocentry}
\includegraphics[]{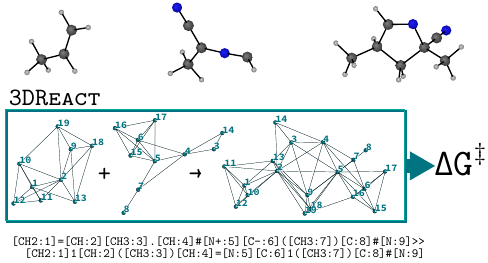}
\end{tocentry}

\end{document}

\title{{\sc Supplementary Information}\texorpdfstring{\\}{}
3DReact: Geometric deep learning for chemical reactions}

\author{Puck van Gerwen}
\affiliation{\LCMD}
\affiliation{\NCCRcat}
\author{Ksenia R. Briling}
\author{Charlotte Bunne}
\affiliation{\NCCRcat}
\affiliation{\LAS}
\author{Vignesh Ram Somnath}
\affiliation{\NCCRcat}
\affiliation{\LAS}
\author{Ruben Laplaza}
\affiliation{\LCMD}
\affiliation{\NCCRcat}
\author{Andreas Krause}
\affiliation{\NCCRcat}
\affiliation{\LAS}
\author{Clemence Corminboeuf}
\email{clemence.corminboeuf@epfl.ch}
\affiliation{\LCMD}
\affiliation{\NCCRcat}

\date{\today}

\maketitle
\onecolumngrid
\tableofcontents
\clearpage

%%%%%%%%%%%%%%%%%%%%%%%%%%%%%%%%%%%%%%%%%%%%%%%%%%%%%%%%%%%%%%%%%%%%%%%%%%%%%%%%%%%%%%%%%%%%%%%%%%%%
\section{Molecular channels}
\label{sec:molecular_channels}
As briefly described in the main text,
a molecule is represented as a distance-based graph where nodes describe atoms and edges
describe bonds. Instead of explicitly using connectivity information, the
``bonds'' of atom $a$ are formed with all the neighboring $\Neigh(a)$ atoms within the
cutoff $r_{\max}$, all the (directed) bonds $\{(a,b)\}$ in the molecule forming
set $\Bonds$.
Initial node (atom) features $\{\vec x^{(0)}_a\}$ encode several cheminformatic
features from \texttt{RDKit},\cite{rdkit2023} including atomic number,
chirality tag (unspecified, tetrahedral, or other, including octahedral, square planar, allene-type, \etc),
number of directly-bonded neighbors, number of rings,
implicit valence, formal charge, number of attached hydrogens, number of radical
electrons, hybridization, aromaticity, presence in rings of specified sizes
from 3 to 7.

Inspired from related models,\cite{corso2023diffdock} initial scalar edge (bond) features $\{\vec e^{(0)}_{ab}\}$ are projections
of the atom distances $|\vec r_{ab}|$ onto $n_g$~Gaussians
uniformly spanning the line segment from $0$ to $r_{\max}$
with the step $\Delta \mu = r_{\max}/(n_g-1)$,
\begin{gather}
\label{eq:initial-edge-features-start}
\vec e^{(0)}_{ab} = \vec f_1(|\vec r_{ab}|)  \qquad \forall (a,b)\in\Bonds,\\
\vec f_1(r)
= \left\{\exp\left(-\frac12\left(\frac{r-n\Delta\mu}{\Delta\mu}\right)^2\right)\right\}
\quad n \in 0, \ldots, n_g{-}1.
\end{gather}
The tensorial edge features $\{\vec z_{ab}\}$, later used as filters, are projections
of normalized difference vectors between atomic positions $\vec r_{ab}/|\vec r_{ab}|$
onto spherical harmonics $Y^\ell_m$ of $0\le\ell\le2$,
\begin{gather}
\vec z_{ab}
\equiv \vec z_{ab}^{0e} \oplus \vec z_{ab}^{1o} \oplus \vec z_{ab}^{2e}
= \vec f_2(\vec r_{ab} / |\vec r_{ab}|)
\qquad \forall (a,b) \in \Bonds,\\
\label{eq:initial-edge-features-end}
\vec f_2(\vec r) = Y^0_0(\vec r)
\oplus\left\{Y^1_m(\vec r)\right\}_{|m| \le 1}
\oplus\left\{Y^2_m(\vec r)\right\}_{|m| \le 2},
\end{gather}
where $\oplus$ denotes the concatenation operator, and the components of $\vec z_{ab}$ are labelled in superscript by the corresponding irreducible representation ($\ell e$ for even parity and $\ell o$ for odd parity) of the $\mathrm{O}(3)$ group.\footnote{\url{https://docs.e3nn.org/en/stable/guide/irreps.html}}
The initial $\vec x^{(0)}$ and $\vec e^{(0)}$ are then passed through embeddings
to give $\vec x^{(1)}_a \ \forall a$ and $\vec e_{ab} \ \forall (a,b)\in\Bonds$.

\begin{figure}[t]
\centering
\includegraphics[width=0.55\textwidth]{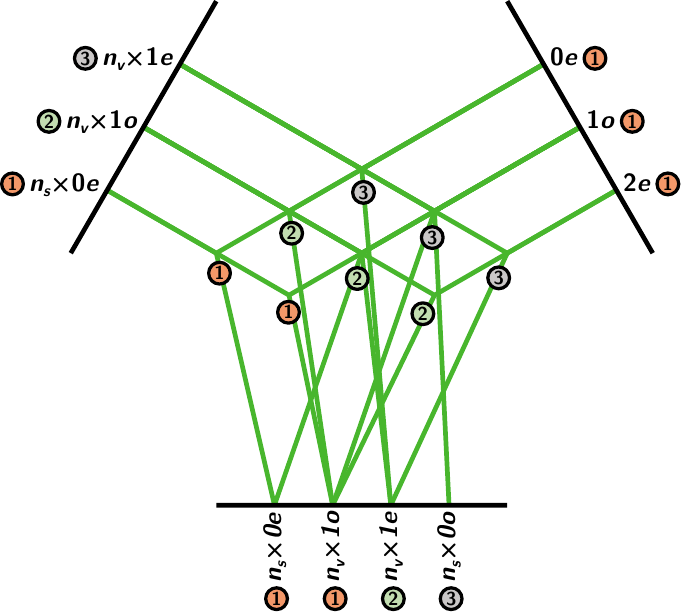}
\caption{Irrep sequence in the (1), (2), (3) convolutional layers of
\textsc{EquiReact}. Input irreps are on the left
(hidden atom and bond features)
and right
(spherical harmonics filters), output irreps are at the
bottom, and paths that connect them are in the middle in green. Note that
formally $2e$ is present in the right input already at the first layer, but
does not contribute to the output.}
\label{fig:irrep_seq}
\end{figure}

The atomic representations $\{\mathbf{x}^{(1)}_a\}$ are updated by $n_{\mathrm{conv}} \in \{2, 3\}$ equivariant convolutional layers:
\begin{DispWithArrows}<\text{\bf Layer 1\:}>
& \label{eq:convs-start}
\:\vec w^{(1)}_{ab}
= \vec g_{31}(\vec e_{ab} \oplus \vec x^{(1)}_{a} \oplus \vec x^{(1)}_{b})
\quad \forall (a,b)\in\Bonds
\\
&\:\begin{aligned}
\vec s^{(1)}_{b}
&\equiv \vec s^{0e(1)}_{b} \oplus \vec s^{1o(1)}_{b}
\\&= \tfrac1{\Neigh(b)} \sum_{\mathrlap{a: (a,b)\in\Bonds}} \vec t_1(\vec x^{(1)}_a, \vec z_{ab}, \vec w^{(1)}_{ab}) \quad\forall b
\end{aligned}
\\&\:\vec x^{0e(2)} = \vec x^{(1)} + \vec s^{0e(1)}
\\&\:\vec {x}^{(2)} = \vec x^{0e(2)} \oplus \vec s^{1o(1)}
\end{DispWithArrows}
\begin{DispWithArrows}<\text{\bf Layer 2\:}>
&\:\vec w^{(2)}_{ab}
= \vec g_{32}(\vec e_{ab} \oplus \vec x^{0e(2)}_{a} \oplus \vec x^{0e(2)}_{b})
\quad \forall (a,b)\in\Bonds
\\
&\:\begin{aligned}
\vec s^{(2)}_{b}
&\equiv \vec s^{0e(2)}_{b} \oplus \vec s^{1o(2)}_{b} \oplus \vec s^{1e(2)}_{b}
\\&= \tfrac1{\Neigh(b)} \sum_{\mathrlap{a: (a,b)\in\Bonds}} \vec t_2(\vec x^{(2)}_a, \vec z_{ab}, \vec w^{(2)}_{ab}) \quad\forall b
\end{aligned}
\\&\:\vec x^{0e(3)} = \vec x^{0e(2)} + \vec s^{0e(2)}
\\&\:\vec {x}^{(3)} = \vec x^{0e(3)} \oplus \big(\vec s^{1o(1)}+\vec s^{1o(2)}\big) \oplus \vec s^{1e(2)}
\end{DispWithArrows}
\begin{DispWithArrows}<\text{\bf Layer 3\:}>
&\:\vec w^{(3)}_{ab}
= \vec g_{33}(\vec e_{ab} \oplus \vec x^{0e(3)}_{a} \oplus \vec x^{0e(3)}_{b})
\quad \forall (a,b)\in\Bonds
\\
&\:\begin{aligned}
\vec s^{(3)}_{b}
& \equiv \vec s^{0e(3)}_{b} \oplus \vec s^{1o(3)}_{b} \oplus \vec s^{1e(3)}_{b} \oplus \vec s^{0o(3)}_{b}
\\& = \tfrac1{\Neigh(b)} \sum_{\mathrlap{a: (a,b)\in\Bonds}} \vec t_3(\vec x^{(3)}_a, \vec z_{ab}, \vec w^{(3)}_{ab}) \quad\forall b
\end{aligned}
\\&\: \vec x^\mathrm{out} = \big(\vec x^{0e(3)} + \vec s^{0e(3)}\big) \oplus \vec s^{0o(3)}.
\end{DispWithArrows}
In \textbf{Layer 1}, for example,
$\vec s^{(1)}_{b}\equiv \vec s^{0e(1)}_{b} \oplus \vec s^{1o(1)}_{b}$
means that the result of the function $\vec t_1$
consists of scalars ($0e$) and vectors ($1o$) that can be treated separately.
Each function $\vec t_n(\vec x, \vec z, \vec w)$ is a fully-connected weighted tensor product,
as defined in \texttt{e3nn},\cite{geiger2020euclidean}
in the form of
\begin{equation}
\label{eq:convs-end}
\vec t_n(\vec x, \vec z, \vec w)
= \bigoplus_k \vec t^{(n)}_k,
\quad \vec t^{(n)}_k = \sum_{uv} w^{(n)}_{uvk} \vec x_u \otimes \vec z_v,
\end{equation}
where $\{k,u,v\}$ index individual tensors.
Note that a tensor here refers to the mathematical object that obeys certain transformation laws, not the multi-dimensional array.
The functions $\{\vec t_n\}$ are specified by signatures of irreducible representations (irreps) of two input and one output $\mathrm{O}(3)$
tensors. The output tensor is a combination of weighted sums of paths
(pairs of input irreps) leading to each output irrep.
The irreducible representation (irrep) sequence in each layer from $1$--$3$
is illustrated in Figure~\ref{fig:irrep_seq},

For example,
irreps of $\vec x^{(0)}_a$ and $\vec z_{ab}$
are $(n_s{\times}0e)$ and $(0e \oplus 1o \oplus 2e)$, respectively,
because the former consists of $n_s$ scalars
and the latter is a direct sum of projections onto spherical harmonics of $\ell=0,1,2$.
The desired output irreps are chosen deliberately
from the possible products of the input irreps and can have any shape,
so the signature of function~$\vec t_1$ is
\begin{equation}
\vec t_1: (n_s{\times}0e)
\otimes (0e \oplus 1o \oplus 2e)
\to (n_s{\times}0e \oplus n_v{\times}1o).
\end{equation}
Thus two paths are created,
\begin{gather}
(n_s{\times}0e) \otimes (1{\times}0e) \to (n_s {\times 0e}) \quad \text{ with } n_s{\times}1{\times}n_s \text{ weights}, \\
(n_s{\times}0e) \otimes (1{\times}1o) \to (n_v {\times 1o}) \quad \text{ with } n_s{\times}1{\times}n_v \text{ weights}.
\end{gather}
The output contains $n_s$ scalars and $n_v$ vectors.
There is one tensor per each bond $(a,b)$,
so a tensor $\vec s^{(1)}_b$ for atom $b$ is an average
of $\vec t_1(\vec x^{(1)}_a, \vec z_{ab}, \vec w^{(1)}_{ab})$ over its neighbors $\{a\}$.
It is used to update $\vec x$
which is convoluted with $\vec z_a$ two more times using functions with signatures
\begin{equation}
\vec t_2: (n_s{\times}0e \oplus n_v{\times}1o)
\otimes (0e \oplus 1o \oplus 2e)
\to (n_s{\times}0e \oplus n_v{\times}1o \oplus n_v{\times}1e),
\end{equation}
adding $n_v$ pseudovectors ($n_s^2 + 2 n_s n_v + 3 n_v^2$ weights total), and
\begin{equation}
\vec t_3: (n_s{\times}0e \oplus n_v{\times}1o \oplus n_v{\times}1e)
\otimes (0e \oplus 1o \oplus 2e)\\
\to (n_s{\times}0e \oplus n_v{\times}1o \oplus n_v{\times}1e \oplus n_s{\times}0o),
\end{equation}
adding $n_s$ pseudoscalars ($n_s^2 + 3 n_s n_v + 6 n_v^2$ weights total).

To obtain the weights $\vec w^{(n)}$ for each convolutional layer $n$,
the spherical parts of $\vec x^{(n)}_a$ and $\vec x^{(n)}_b$
are concatenated with the bond features $\vec e_{ab}$
and passed through a multi-layer perceptron $\vec g_{3n}$.

The output of the equivariant molecular channels
is the local molecular representation $\vec X\in \Re^{N_{\mathrm{at}} \times D}$
corresponding to $N_{\mathrm{at}}$ atoms associated with $D$ features.
Depending on the \texttt{sum\_mode} hyperparameter,
it is constructed either from the node features $\{\vec x^\mathrm{out}_a\}$ (\texttt{node} mode)
or both node and edge features $\{\vec x^\mathrm{out}_a \oplus \sum_{b:(a,b)\in\Bonds}\vec e^{(0)}_{ab}\}$ (\texttt{both} mode).
In the case of $n_{\mathrm{conv}}=2$, the vectors $\{\vec x^{0e(3)}_a\}$ are taken to construct the molecular representation.

Note that even through the resulting node features $\vec X$ should be invariant
(with $n_{\mathrm{conv}}=2$) or equivariant (with $n_{\mathrm{conv}}=3$) with coordinate inversion,
the chirality tag, if present in the SMILES, changes the initial node features $\{\vec x^{(0)}_a\}$
and may lead to different representations for two enantiomers.

\bigskip

Above is described the architecture of \textsc{EquiReact}, the equivariant option of \textsc{3DReact}.
The invariant option (\textsc{InReact}) is implemented as a simplified version
of the latter: the spherical harmonics filters take only $\ell=0$
(hence $\vec z_{ab} = 1 \ \forall (a,b) \in \Bonds$),
the convolutional layers signatures become
\begin{equation}
\vec t'_n: (n_s{\times}0e) \otimes (0e) \to (n_s{\times}0e), \quad n\in\{1,2,3\},
\end{equation}
and Eq.~\ref{eq:convs-end} is simplified to a dot product
\begin{equation}
\vec t'_n(\vec x, \vec z, \vec w)
\equiv \vec t'_n(\vec x, \vec w)
= \bigoplus_k \vec t^{\prime(n)}_k,
\quad \vec t^{(n)}_k = \sum_{u} w^{(n)}_{uk} x_u.
\end{equation}

%%%%%%%%%%%%%%%%%%%%%%%%%%%%%%%%%%%%%%%%%%%%%%%%%%%%%%%%%%%%%%%%%%%%%%%%%%%%%%%%%%%%%%%%%%%%%%%%%%%%
\newpage
\section{Model hyperparameters}
\label{sec:model-params}
The best model hyperparameters obtained after the sweep on \textsc{EquiReact} are summarized in Table~\ref{tab:model-params}.
These are the hyperparameters used in all \textsc{3DReact} models in the main text.
\begin{table}[h!]
\begin{tabular}{@{}c@{\hskip 1.5em}c@{\hskip 1.5em}c@{\hskip 1.5em}c@{}} \toprule
Parameter                    & \gdb            & \cyclo          & \proparg          \\\midrule
$n_s$                        & 64              & 64              & 64                \\% n_s
$n_v$                        & 64              & 48              & 64                \\% n_v
$n_g$                        & 32              & 48              & 64                \\% distance_emb_dim
$n_\mathrm{conv}$            & 2               & 2               & 3                 \\% n_conv_layers
$r_{\max}$, \AA              & 2.5             & 2.5             & 5                 \\% radius
$n_{\mathrm{neigh}}$         & 10              & 50              & 10                \\% max_neighbors
$p_d$                        & 0.05            & 0.1             & 0.05              \\% dropout_p
\texttt{sum\_mode}           & \texttt{both}   & \texttt{node}   & \texttt{node}     \\% sum_mode
\texttt{combine\_mode}       & \texttt{diff}   & \texttt{diff}   & \texttt{diff}     \\% combine_mode
\texttt{graph\_mode}         & \texttt{vector} & \texttt{energy} & \texttt{vector}   \\% graph_mode
learning rate                & \num{5e-4}      & \num{e-3}       & \num{e-3}         \\% lr
weight decay                 & \num{e-5}       & \num{e-5}       & \num{e-5}         \\% weight_decay
\bottomrule
\end{tabular}
\caption{Best model hyperparameters as a result of the sweeps.}
\label{tab:model-params}
\end{table}

%%%%%%%%%%%%%%%%%%%%%%%%%%%%%%%%%
\clearpage
\section{Root-mean-square errors}
\label{sec:rmse}
The results presented in Table~\extref{tab:model_performance_all}
and Table~\extref{tab:model_performance_equireact}
are repeated in Table~\ref{tab:model_performance_all_rmse} using root mean squared errors (RMSEs) rather than mean absolute errors (MAEs)
as the performance metric.
\begin{table}[h]
\begin{tabular}{@{}cccccc@{}} \toprule
\makecell{Dataset \\ (property, units)} & \makecell{Atom-mapping\\ regime}
            & \CGR & SLATM$_d$+KRR & \textsc{InReact} & \textsc{EquiReact} \\
\midrule
\multicolumn{6}{@{}c@{}}{\emph{Random splits}}\\ \midrule
\multirow{3}{*}{\makecell{\gdb \\ ($\Delta E^\ddag$, kcal/mol)}}
& True & $ 7.6 \pm 0.3 $ & --- & $ 8.4 \pm 0.4 $ & $ 8.4 \pm 0.3 $ \\
& RXNMapper & $ 9.57 \pm 0.28 $ & --- & $ 9.9 \pm 0.4 $ & $ 10.0 \pm 0.4 $ \\
& None & $ 13.2 \pm 0.5 $ & $ 10.8 \pm 0.4 $ & $ 10.5 \pm 0.4 $ & $ 10.4 \pm 0.5 $ \\
\\[0.002cm]\multirow{3}{*}{\makecell{\cyclo \\ ($\Delta G^\ddag$, kcal/mol)}}
& True & $ 3.70 \pm 0.20 $ & --- & $ 3.33 \pm 0.18 $ & $ 3.24 \pm 0.21 $ \\
& RXNMapper & $ 3.72 \pm 0.16 $ & --- & $ 3.31 \pm 0.21 $ & $ 3.29 \pm 0.26 $ \\
& None & $ 3.73 \pm 0.23 $ & $ 3.65 \pm 0.22 $ & $ 3.34 \pm 0.18 $ & $ 3.24 \pm 0.23 $ \\
\\[0.002cm] \multirow{2}{*}{\makecell{\proparg \\ ($\Delta E^\ddag$, kcal/mol)}}
& True & $ 1.97 \pm 0.16 $ & --- & $ 0.58 \pm 0.16 $ & $ 0.52 \pm 0.14 $ \\
& None & $ 2.01 \pm 0.19 $ & $ 0.52 \pm 0.09 $ & $ 0.60 \pm 0.16 $ & $ 0.49 \pm 0.09 $ \\
\midrule
\multicolumn{6}{@{}c@{}}{\emph{Scaffold splits}}\\ \midrule
\multirow{3}{*}{\makecell{\gdb \\ ($\Delta E^\ddag$, kcal/mol)}}
& True & $ 10.6 \pm 0.9 $ & --- & $ 11.6 \pm 1.0 $ & $ 11.5 \pm 1.2 $ \\
& RXNMapper & $ 12.8 \pm 0.9 $ & --- & $ 13.3 \pm 1.0 $ & $ 13.3 \pm 1.1 $ \\
& None & $ 17.4 \pm 0.8 $ & $ 15.2 \pm 1.3 $ & $ 14.5 \pm 1.1 $ & $ 14.4 \pm 1.1 $ \\
\\[0.002cm]\multirow{3}{*}{\makecell{\cyclo \\ ($\Delta G^\ddag$, kcal/mol)}}
& True & $ 3.9 \pm 0.4 $ & --- & $ 3.7 \pm 0.4 $ & $ 3.7 \pm 0.4 $ \\
& RXNMapper & $ 3.9 \pm 0.4 $ & --- & $ 3.7 \pm 0.4 $ & $ 3.7 \pm 0.5 $ \\
& None & $ 4.1 \pm 0.4 $ & $ 4.0 \pm 0.4 $ & $ 3.7 \pm 0.4 $ & $ 3.7 \pm 0.4 $ \\
\\[0.002cm] \multirow{2}{*}{\makecell{\proparg \\ ($\Delta E^\ddag$, kcal/mol)}}
& True & $ 2.2 \pm 0.3 $ & --- & $ 0.71 \pm 0.22 $ & $ 0.65 \pm 0.17 $ \\
& None & $ 2.2 \pm 0.4 $ & $ 0.62 \pm 0.16 $ & $ 0.68 \pm 0.18 $ & $ 0.64 \pm 0.19 $ \\
\bottomrule
\end{tabular}
\caption{
Performance as measured in root-mean-square errors (RMSEs) of predictions
of \textsc{3DReact} (\textsc{InReact} and \textsc{EquiReact}) \vs \CGR and SLATM$_d$.
\textsc{3DReact}$_M$ is used for the ``True'' and ``RXNMapper'' regimes,
and \textsc{3DReact}$_S$ is used for the ``None'' regime.
RMSEs are averaged over 10~folds of random/scaffold 80/10/10 splits (training/validation/test)
and reported together with standard deviations across folds.
}
\label{tab:model_performance_all_rmse}
\end{table}

%%%%%%%%%%%%%%%%%%%%%%%%%%%%%%%%%%%%%%%%%%%%%%%%%%%%%%%%%%%%%%%%%%%%%%%%%%%%%%%%%%%%%%%%%%%%%%%%%%%%
\clearpage
\section{Cross-attention as a surrogate for atom-mapping in the ``None'' regime}
\label{sec:cross}

Since the ``True'' atom-mapping regime allowed for a significantly improved model
over the ``RXNMapper'' and ``None'' regimes for the \gdb dataset,
we thought that a model based on cross-attention between reactants and products (\textsc{3DReact}$_X$)
could provide a competitive surrogate to the atom-mapping-based models.

Given queries $\vec Q \in \Re^{N \times D}$,
keys $\vec K \in \Re^{M \times D}$
and values $\vec V \in \Re^{M \times D}$,
attention is computed as
\begin{equation}
\label{eq:attention}
\vec{A} = \softmax \left( \frac{ \vec{Q} \vec{K}^T}{\sqrt{D}} \right)
\end{equation}
and the ``reordered'' values $\vec Y$ are
\begin{equation}
\label{eq:reorder}
\vec{Y} = \vec A \vec{V}.
\end{equation}
We used the implementation of this scaled-dot-product attention\cite{vaswani2017attention} in
\texttt{PyTorch}'s\cite{paszke2019pytorch} \texttt{MultiheadAttention} (\texttt{PyTorch} version 1.12.1).
The representations are re-ordered using Eq.~\ref{eq:attention} and Eq.~\ref{eq:reorder} with $\vec{Q}$
as the vector representation of reactants, $\vec{K}$ and $\vec{V}$ as the vector representations of products,
and vice versa (thus here $N=M=N_\mathrm{at}$).

However, we found that
the cross-attention module
does not improve over
the simple model \textsc{3DReact}$_S$,
based only on geometries of isolated reactants and products
without any information exchange between them,
for all three datasets (Table~\ref{tab:cross-attention}).

\begin{table}[h!]
\begin{tabular}{@{}ccccc@{}} \toprule
Dataset (property, units)
& \textsc{InReact}$_X$ & \textsc{InReact}$_S$
& \textsc{EquiReact}$_X$ & \textsc{EquiReact}$_S$
\\ \midrule
\gdb ($\Delta E^\ddag$, kcal/mol)& $ 6.68 \pm 0.27 $& $ 6.56 \pm 0.26 $& $ 6.7 \pm 0.3 $& $ 6.53 \pm 0.28 $ \\[0.002cm]
\cyclo ($\Delta G^\ddag$, kcal/mol)& $ 2.53 \pm 0.09 $& $ 2.39 \pm 0.05 $& $ 2.43 \pm 0.09 $& $ 2.31 \pm 0.09 $ \\[0.002cm]
\proparg ($\Delta E^\ddag$, kcal/mol)& $ 0.34 \pm 0.05 $& $ 0.34 \pm 0.06 $& $ 0.33 \pm 0.06 $& $ 0.31 \pm 0.06 $ \\[0.002cm]
\bottomrule
\end{tabular}
\caption{Performance of the alternative models \textsc{3DReact}$_X$ and \textsc{3DReact}$_S$ in the None mapping mode.}
\label{tab:cross-attention}
\end{table}

Investigating further for the \gdb dataset, we wanted to find out whether the attention module
could infer atom-mapping from an easier, supervised learning scenario.
The \textsc{Mapper} model took the graphs of reactants and products as input,
as for \textsc{3DReact}.
Since the atom indices in the reactants were ordered sequentially
($1, 2, 3, \dots$), the objective was to learn the permutation of the atom
indices in the products, in order to map them to the reactants correctly. The
true atom maps provided by the \gdb set\cite{gdb7-22-ts} were used to train and
validate the model. The permutation was learned using cross-attention
between atoms in reactants and products (Eq.~\ref{eq:attention})
using queries $\vec Q$ from the vector representation of reactants
and keys $\vec K$ from vector representations of products.

The model was trained using the Adam optimizer with default parameters. The hyperparameters used for the \textsc{3DReact} (\textsc{EquiReact}) components of the \textsc{Mapper} model (to construct the molecular representations) are summarized in Table~\ref{tab:hypers-mapper}.

\begin{table}[h!]
\begin{tabular}{@{}cccccccc@{}} \toprule
Parameter & $n_s$ & $n_v$ & $n_g$ & $n_{\mathrm{conv}}$ & $r_{\mathrm{max}}$ & $n_{\mathrm{neigh}}$ & $p_d$\\ \midrule
Value & 16 & 16 & 32 & 2 & 10 & 20 & 0.1 \\
\bottomrule
\end{tabular}
\caption{Hyperparameters used to construct molecular components in the \textsc{Mapper} model.}
\label{tab:hypers-mapper}
\end{table}

\begin{figure}[t]
\centering
\includegraphics[width=0.8\textwidth]{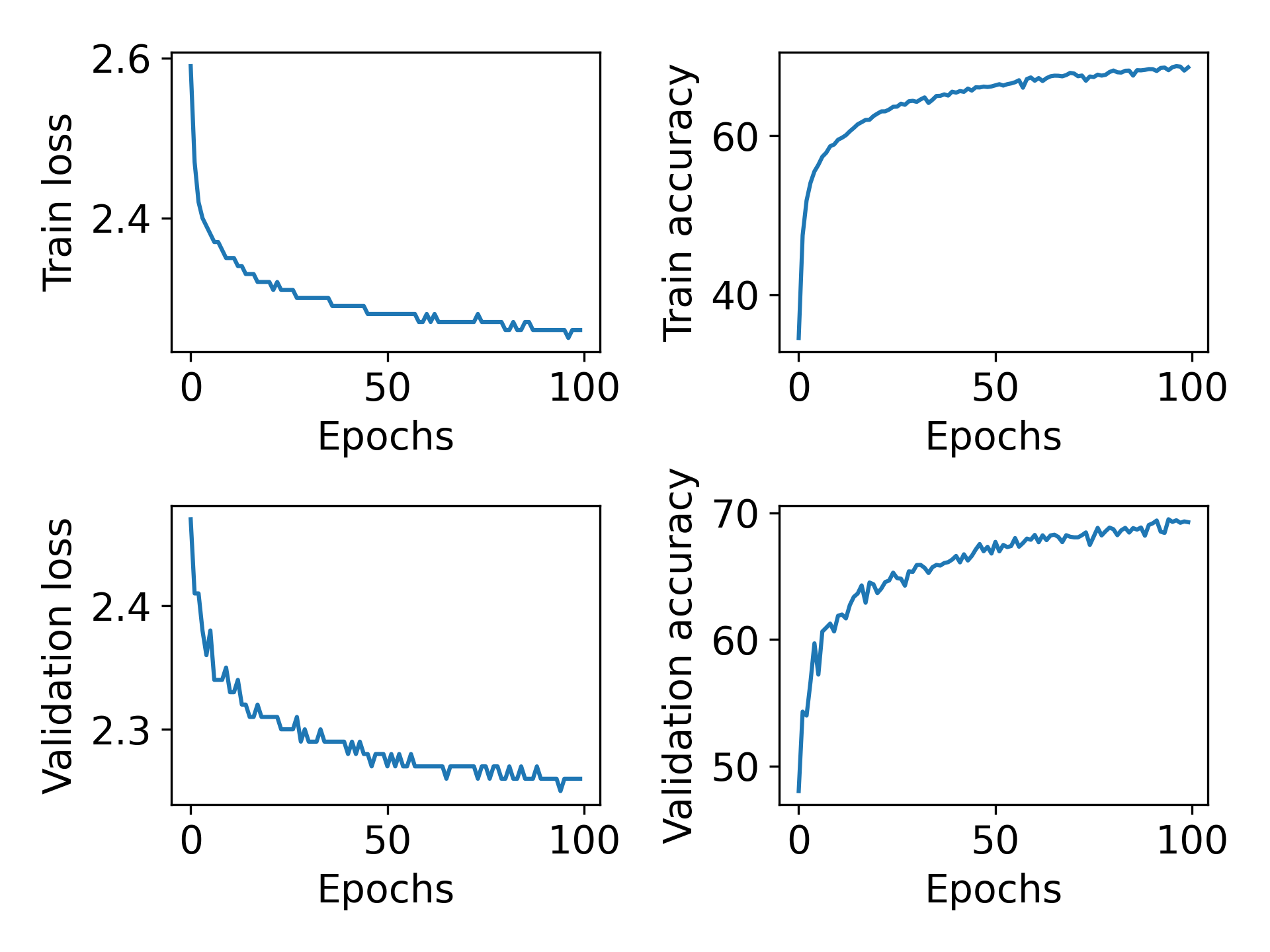}
\caption{
Evolution of the training loss, training accuracy (\%), validation
loss, validation  accuracy (\%) for the \textsc{Mapper} model, trained to
learn atom-mapping in a supervised setting.
}
\label{fig:training-mapper}
\end{figure}

The
optimization objective was to minimize the cross-entropy loss between the true
mapping permutation and the permutation learned by the model. \textsc{Mapper} was run
for 100~epochs on a single random 80/10/10 split of the \gdb dataset. The resulting
training and validation curves are shown in Figure~\ref{fig:training-mapper}.

While the model learns, both
the training and validation accuracy saturate at around 70\%,
suggesting there is an upper bound to the accuracy of the \textsc{Mapper} model.
This may be related to the unconstrained nature of the optimization:
while the predicted maps can converge to any integer value,
each integer prediction should be taken from the pool of integers (atom-map labels)
in the reactants, and appear only once.
The architecture is also much simpler than that of \textsc{RXNMapper}\cite{schwaller2021extraction} for example.

Nevertheless, the fact that the model is able to learn atom-mapping
to a reasonable degree of accuracy suggests that the cross-attention module in
\textsc{3DReact} could mimic a low-quality atom-mapping.
Therefore, \textsc{3DReact}$_X$ likely fails to improve model performance
because the unsupervised atom-mapping task is embedded in a broader learning context.
It seems that unlike the supervised task of reaction outcome prediction,\cite{schwaller2021extraction}
a graph-based model trained to predict reaction barriers does not learn chemical signatures
like atom-mapping as part of the training process.
Other information (like that based on the geometries of reactants and products)
is instead exploited to make good predictions.

%%%%%%%%%%%%%%%%%%%%%%%%%%%%%%%%%%%%%%%%%%%%%%%%%%%%%%%%%%%%%%%%%%%%%%%%%%%%%%%%%%%%%%%%%%%%%%%%%%%%
\section{Extrapolation studies}
\label{sec:extrapolation}
Table~\ref{tab:model_performance_all_rmse} shows the results for an extensive set of extrapolation studies, both for \textsc{InReact} and \textsc{EquiReact} as well as for the baseline models.
Property-based splits order the training/validation/test sets
according to the target value in ascending or descending order.
The nonzero standard deviations for neural network models
arise from different organization of the datapoints into batches.
Size-based splits instead order the datapoints into training/validation/test sets according to molecular size,
\ie, number of heavy atoms
(reactant size for \gdb and \proparg, product size for \cyclo), again sorted in ascending or descending order.
Since the molecular size is a discrete variable, the initial data shuffling
affects the composition of the sets and leads to the non-zero standard deviations for all the models.

In the size-based splits, ascending is more challenging than descending, since a model trained on large molecules also has a notion of atoms in smaller molecules. In principle, property-based splits are equally challenging as ascending or descending, but due to asymmetry in the data distributions, ascending seems to result in higher test MAEs for all models.

\begin{table}[p]
\begin{tabular}{@{}cccccc@{}} \toprule
\small
\makecell{Dataset \\ (property, units)} & \makecell{Atom-mapping regime}
            & \CGR & SLATM$_d$+KRR & \textsc{InReact} & \textsc{EquiReact} \\
\midrule
\multicolumn{6}{@{}c@{}}{\emph{Scaffold splits}}\\ \midrule
\multirow{3}{*}{\makecell{\gdb \\ ($\Delta E^\ddag$, kcal/mol)}}
& True      & $ 7.1 \pm 0.6 $  & ---              & $ 7.8 \pm 0.7 $  & $ 7.8 \pm 0.8 $ \\
& RXNMapper & $ 8.7 \pm 0.8 $  & ---              & $ 9.2 \pm 0.8 $  & $ 9.1 \pm 0.8 $ \\
& None      & $ 12.9 \pm 0.6 $ & $\bf 11.0 \pm 1.0 $ & $\bf  10.1 \pm 0.9 $ & $\bf  10.0 \pm 0.9 $ \\
\\[-1ex]\multirow{3}{*}{\makecell{\cyclo \\ ($\Delta G^\ddag$, kcal/mol)}}
& True      & $ 2.92 \pm 0.24 $ & ---               & $ 2.79 \pm 0.18 $ & $ 2.72 \pm 0.18 $ \\
& RXNMapper & $ 2.92 \pm 0.23 $ & ---               & $ 2.77 \pm 0.22 $ & $ 2.71 \pm 0.23 $ \\
& None      & $ 3.1 \pm 0.3 $   & $ 2.97 \pm 0.22 $ & $ 2.76 \pm 0.22 $ & $ 2.72 \pm 0.19 $ \\
\\[-1ex] \multirow{2}{*}{\makecell{\proparg \\ ($\Delta E^\ddag$, kcal/mol)}}
& True & $ 1.64 \pm 0.21 $ & ---               & $\bf  0.44 \pm 0.11 $ & $\bf  0.40 \pm 0.08 $ \\
& None & $ 1.64 \pm 0.19 $ & $\bf  0.36 \pm 0.05 $ & $\bf  0.45 \pm 0.10 $ & $\bf  0.41 \pm 0.09 $ \\
\midrule
\multicolumn{6}{@{}c@{}}{\emph{Property-based splits (ascending)}}\\ \midrule
\multirow{3}{*}{\makecell{\gdb \\ ($\Delta E^\ddag$, kcal/mol)}}
& True      & $\bf  29.0 \pm 0.9 $ & ---       & $ 30.9 \pm 0.6 $ & $ 30.9 \pm 0.5 $ \\
& RXNMapper & $ 31.7 \pm 0.9 $ & ---       & $ 33.0 \pm 0.4 $ & $ 32.2 \pm 0.6 $ \\
& None      & $ 36.0 \pm 0.4 $ & $\bf  34.80 $ & $ 35.3 \pm 0.3 $ & $ 35.3 \pm 0.4 $ \\
\\[-1ex]\multirow{3}{*}{\makecell{\cyclo \\ ($\Delta G^\ddag$, kcal/mol)}}
& True      & $ 11.7 \pm 0.3 $ & ---       & $\bf  9.0 \pm 0.6 $ & $\bf  9.3 \pm 0.4 $ \\
& RXNMapper & $ 12.1 \pm 0.3 $ & ---       & $\bf  8.8 \pm 0.7 $ & $\bf  8.8 \pm 0.6 $ \\
& None      & $ 11.9 \pm 0.3 $ & $ 12.14 $ & $\bf  9.5 \pm 0.3 $ & $\bf  9.6 \pm 0.4 $ \\
\\[-1ex] \multirow{2}{*}{\makecell{\proparg \\ ($\Delta E^\ddag$, kcal/mol)}}
& True & $ 5.10 \pm 0.12 $ & ---      & $\bf  4.31 \pm 0.15 $ & $\bf  4.08 \pm 0.11 $ \\
& None & $ 5.09 \pm 0.13 $ & $ 5.83 $ & $\bf  4.53 \pm 0.16 $ & $\bf  4.33 \pm 0.14 $ \\
\midrule
\multicolumn{6}{@{}c@{}}{\emph{Property-based splits (descending)}}\\ \midrule
\multirow{3}{*}{\makecell{\gdb \\ ($\Delta E^\ddag$, kcal/mol)}}
& True      & $ 24.0 \pm 1.1 $ & ---       & $ 24.9 \pm 0.9 $ & $ 25.0 \pm 0.7 $ \\
& RXNMapper & $ 25.7 \pm 0.7 $ & ---       & $ 26.8 \pm 0.9 $ & $ 26.4 \pm 0.4 $ \\
& None      & $ 30.3 \pm 0.5 $ & $ 31.08 $ & $ 30.6 \pm 0.6 $ & $ 31.0 \pm 0.3 $ \\
\\[-1ex]\multirow{3}{*}{\makecell{\cyclo \\ ($\Delta G^\ddag$, kcal/mol)}}
& True      & $ 6.1 \pm 0.4 $   & ---      & $\bf  5.03 \pm 0.28 $ & $\bf  4.8 \pm 0.5 $ \\
& RXNMapper & $ 5.9 \pm 0.8 $   & ---      & $ 5.2 \pm 0.4 $   & $ 4.6 \pm 0.5 $ \\
& None      & $ 6.22 \pm 0.30 $ & $ 6.46 $ & $\bf  5.18 \pm 0.17 $ & $\bf  5.07 \pm 0.28 $ \\
\\[-1ex] \multirow{2}{*}{\makecell{\proparg \\ ($\Delta E^\ddag$, kcal/mol)}}
& True & $ 3.33 \pm 0.20 $ & ---      & $\bf  1.58 \pm 0.12 $ & $ \bf 1.69 \pm 0.20 $ \\
& None & $ 3.28 \pm 0.23 $ & $ 2.02 $ & $\bf  1.70 \pm 0.11 $ & $ 2.13 \pm 0.27 $ \\
\midrule
\multicolumn{6}{@{}c@{}}{\emph{Size-based splits (ascending)}}\\ \midrule
\multirow{3}{*}{\makecell{\gdb \\ ($\Delta E^\ddag$, kcal/mol)}}
& True      & $\bf  5.62 \pm 0.20 $  & ---               & $ 6.41 \pm 0.25 $ & $ 6.36 \pm 0.24 $ \\
& RXNMapper & $\bf  6.97 \pm 0.25 $  & ---               & $ 7.52 \pm 0.19 $ & $ 7.55 \pm 0.28 $ \\
& None      & $ 10.39 \pm 0.14 $ & $\bf  8.06 \pm 0.16 $ & $\bf  8.10 \pm 0.22 $ & $\bf  8.11 \pm 0.23 $ \\
\\[-1ex]\multirow{3}{*}{\makecell{\cyclo \\ ($\Delta G^\ddag$, kcal/mol)}}
& True      & $ 4.57 \pm 0.15 $ & ---               & $\bf 3.86 \pm 0.13 $ & $\bf 3.86 \pm 0.11 $ \\
& RXNMapper & $ 4.68 \pm 0.13 $ & ---               & $\bf 3.91 \pm 0.13 $ & $\bf 3.89 \pm 0.21 $ \\
& None      & $ 4.57 \pm 0.11 $ & $ 4.79 \pm 0.03 $ & $\bf 3.96 \pm 0.13 $ & $\bf 3.96 \pm 0.16 $ \\
\\[-1ex] \multirow{2}{*}{\makecell{\proparg \\ ($\Delta E^\ddag$, kcal/mol)}}
& True & $ 1.537 \pm 0.023 $ & ---                 & $ \bf 0.52 \pm 0.07 $ & $ \bf 0.48 \pm 0.05 $ \\
& None & $ 1.538 \pm 0.026 $ & $ \bf 0.504 \pm 0.021 $ & $ 0.63 \pm 0.06 $ & $ \bf 0.55 \pm 0.05 $ \\
\midrule
\multicolumn{6}{@{}c@{}}{\emph{Size-based splits (descending)}}\\ \midrule
\multirow{3}{*}{\makecell{\gdb \\ ($\Delta E^\ddag$, kcal/mol)}}
& True      & $ \bf 3.25 \pm 0.05 $ & ---               & $ 3.78 \pm 0.14 $ & $ 3.74 \pm 0.10 $ \\
& RXNMapper & $ 4.61 \pm 0.15 $ & ---               & $ 4.75 \pm 0.12 $ & $ 4.74 \pm 0.13 $ \\
& None      & $ 8.36 \pm 0.22 $ & $ \bf 4.85 \pm 0.07 $ & $ \bf 4.90 \pm 0.11 $ & $ \bf 4.90 \pm 0.11 $ \\
\\[-1ex]\multirow{3}{*}{\makecell{\cyclo \\ ($\Delta G^\ddag$, kcal/mol)}}
& True      & $ 2.85 \pm 0.08 $ & ---               & $ 2.74 \pm 0.10 $ & $ 2.72 \pm 0.06 $ \\
& RXNMapper & $ 2.88 \pm 0.06 $ & ---               & $ 2.77 \pm 0.09 $ & $ 2.72 \pm 0.07 $ \\
& None      & $ 2.85 \pm 0.07 $ & $ 2.80 \pm 0.03 $ & $ 2.85 \pm 0.07 $ & $ 2.76 \pm 0.09 $ \\
\\[-1ex] \multirow{2}{*}{\makecell{\proparg \\ ($\Delta E^\ddag$, kcal/mol)}}
& True & $ 1.30 \pm 0.04 $ & ---                 & $ \bf 0.278 \pm 0.021 $ & $ \bf 0.263 \pm 0.023 $ \\
& None & $ 1.30 \pm 0.04 $ & $ 0.307 \pm 0.011 $ & $ \bf 0.236 \pm 0.029 $ & $ \bf 0.226 \pm 0.028 $ \\
\bottomrule
\end{tabular}
\caption{
Performance as measured in mean absolute errors (MAEs) of predictions
of \textsc{3DReact} (\textsc{InReact} and \textsc{EquiReact}) \vs \CGR and SLATM$_d$.
MAEs are averaged over 10~folds of 80/10/10 splits (training/validation/test)
and reported together with standard deviations across folds.
Lowest errors are highlighted in bold, if there are statistically meaningful differences between models in each regime/dataset/split type tested.
}
\label{tab:model_performance_all_splits}
\end{table}

%%%%%%%%%%%%%%%%%%%%%%%%%%%%%%%%%%%%%%%%%%%%%%%%%%%%%%%%%%%%%%%%%%%%%%%%%%%%%%%%%%%%%%%%%%%%%%%%%%%%
\clearpage
\section{Correlation plots for the \gdb dataset}
\label{sec:gdb_outliers_corr}
\begin{figure}[t]
\centering
\includegraphics[width=\textwidth]{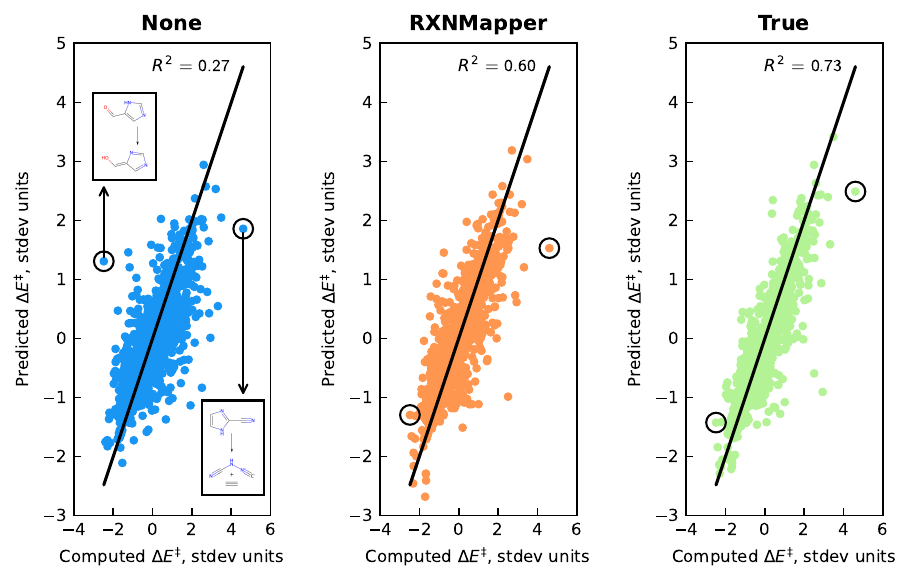}
\caption{
Correlation plots of predicted with \textsc{3DReact} (\textsc{InReact}) $\Delta E^\ddag$ values \vs true (computed) labels
for the first reactant-based scaffold split on the \gdb dataset.
The same test reactions are highlighted in each subplot.
}
\label{fig:gdb_outliers_corr}
\end{figure}

Figure~\ref{fig:gdb_outliers_corr} illustrates that for the same scaffold split, from ``None'' to ``RXNMapper'' to ``True'' the coefficient of determination ($R^2$) increases and outliers successively move closer to the $y=x$ line.

%%%%%%%%%%%%%%%%%%%%%%%%%%%%%%%%%%%%%%%%%%%%%%%%%%%%%%%%%%%%%%%%%%%%%%%%%%%%%%%%%%%%%%%%%%%%%%%%%%%%

%%%%%%%%%%%%%%%%%%%%%%%%%%%%%%%%%%%%%%%%%%%%%%%%%%%%%%%%%%%%%%%%%%%%%%%%%%%%%%%%%%%%%%%%%%%%%%%%%%%%
\section{Performance with and without explicit hydrogen atoms}
\label{sec:hydrogens}

The results presented in the main text were for models built from molecular graphs constructed without
explicit inclusion of hydrogen atoms as nodes.
The performance for models including H-nodes is shown in Table~\ref{tab:models-with-withoutH}.

For the \gdb dataset, the best-performing model in the ``True'' regime (\textsc{ChemProp}) is with explicit Hs. Since this set includes \ce{H2}-abstraction reactions, the explicit inclusion of hydrogen in combination with the reaction mechanism in the form of atom-mapping is particularly informative. Interestingly, \textsc{3DReact} ``True'' and ``None`` does not suffer when removing H atoms. In the ``RXNMapper'' regime, the inclusion of H-nodes increases the errors especially for \textsc{ChemProp}.
The atom-mapping provided by \textsc{RXNMapper} is inferred as a consequence of a related task: predicting randomly masked parts of a reaction sequence (reaction SMILES).\cite{schwaller2021extraction} Since there are few H atoms in reaction SMILES strings (typically they are implicit), it is natural that \textsc{RXNMapper} struggles to map H atoms in reactions.

For the \cyclo and \proparg datasets, no models benefit from the inclusion of explicit Hs in any regime.
Since larger molecules participate in these reactions
which never consist purely of \ce{H2}-abstraction, the reactions are well-described without an explicit description of H atoms.
The cost of the message passing increases considerably when including H-nodes, and \textsc{3DReact}'s performance is not strongly correlated with their inclusion. This coupled with the fact that atom-mapping tools usually do not map hydrogens, as well as the fact that most atom-mapped reaction SMILES have maps only for heavy atoms, resulted in the decision that we use the models without explicit H-nodes in \textsc{ChemProp} and \textsc{3DReact}.

\begin{table}[ht!]
\begin{tabular}{@{}cccccccc@{}} \toprule
\multirow{3}{*}{\makecell{Dataset \\ (property, units)}}&
\multirow{3}{*}{H mode}&
\multicolumn{6}{c}{Atom mapping regime} \\ \cmidrule(lr){3-8}
&& \multicolumn{2}{@{}c@{}}{True} & \multicolumn{2}{@{}c@{}}{RXNMapper} & \multicolumn{2}{@{}c@{}}{None} \\ \cmidrule(lr){3-4} \cmidrule(lr){5-6}  \cmidrule(lr){7-8}
&& \CGR & \textsc{3DReact}$_M$ & \CGR & \textsc{3DReact}$_M$ & \CGR & \textsc{3DReact}$_S$ \\
\midrule \multirow{2}{*}{\makecell{\gdb \\ ($\Delta E^\ddag$, kcal/mol)}}
& with  & $ \mathbf{4.12 \pm 0.13} $ & $ 4.90 \pm 0.16 $ & $ 6.36 \pm 0.09 $ & $ 6.24 \pm 0.21 $ & $ 8.87 \pm 0.28 $ & $ \mathbf{6.54 \pm 0.25} $\\
& w/o  & $ 4.35 \pm 0.15 $ & $ 4.93 \pm 0.18 $ & $ \mathbf{5.69 \pm 0.17} $ & $ 6.03 \pm 0.26 $ & $ 9.04 \pm 0.21 $ & $ \mathbf{6.56 \pm 0.26} $\\
\\[0.002cm] \multirow{2}{*}{\makecell{\cyclo \\ ($\Delta G^\ddag$, kcal/mol)}}
& with  & --- & $ \mathbf{2.33 \pm 0.07} $ & $ 2.79 \pm 0.12 $ & $ \mathbf{2.37 \pm 0.07} $ & $ 2.76 \pm 0.10 $ & $ \mathbf{2.38 \pm 0.08} $\\
& w/o  & $ 2.69 \pm 0.10 $ & $ \mathbf{2.39 \pm 0.08} $ & $ 2.71 \pm 0.07 $ & $ \mathbf{2.37 \pm 0.07} $ & $ 2.71 \pm 0.12 $ & $ \mathbf{2.39 \pm 0.05} $\\
\\[0.002cm] \multirow{2}{*}{\makecell{\proparg \\ ($\Delta E^\ddag$, kcal/mol)}}
& with  & $ 1.55 \pm 0.16 $ & $ \mathbf{0.38 \pm 0.07} $ & --- & --- & $ 1.54 \pm 0.14 $ & $ \mathbf{0.37 \pm 0.05} $\\
& w/o  & $ 1.53 \pm 0.14 $ & $ \mathbf{0.33 \pm 0.07} $ & --- & --- & $ 1.56 \pm 0.16 $ & $ \mathbf{0.34 \pm 0.06} $\\
\bottomrule
\end{tabular}
\caption{
Performance of \textsc{3DReact} (\textsc{InReact})
with explicit hydrogens as nodes in the graphs (H mode ``with'')
and without
(H mode ``w/o'', as in the main text).
\cyclo ``True'' with Hs is missing because the dataset provides atom maps for heavy atoms only,
while \textsc{3DReact} does not suffer from that since it uses atom-mapped xyz files
(see Section~\extref{sec:methods:datasets}).
MAEs are averaged over 10~folds of 80/10/10 splits (training/validation/test)
and reported together with standard deviations across folds.
Lowest errors, if statistically relevant, in each dataset/mapping regime are highlighted in bold.
}
\label{tab:models-with-withoutH}
\end{table}

\begin{figure}
\centering
\includegraphics[width=0.666\linewidth]{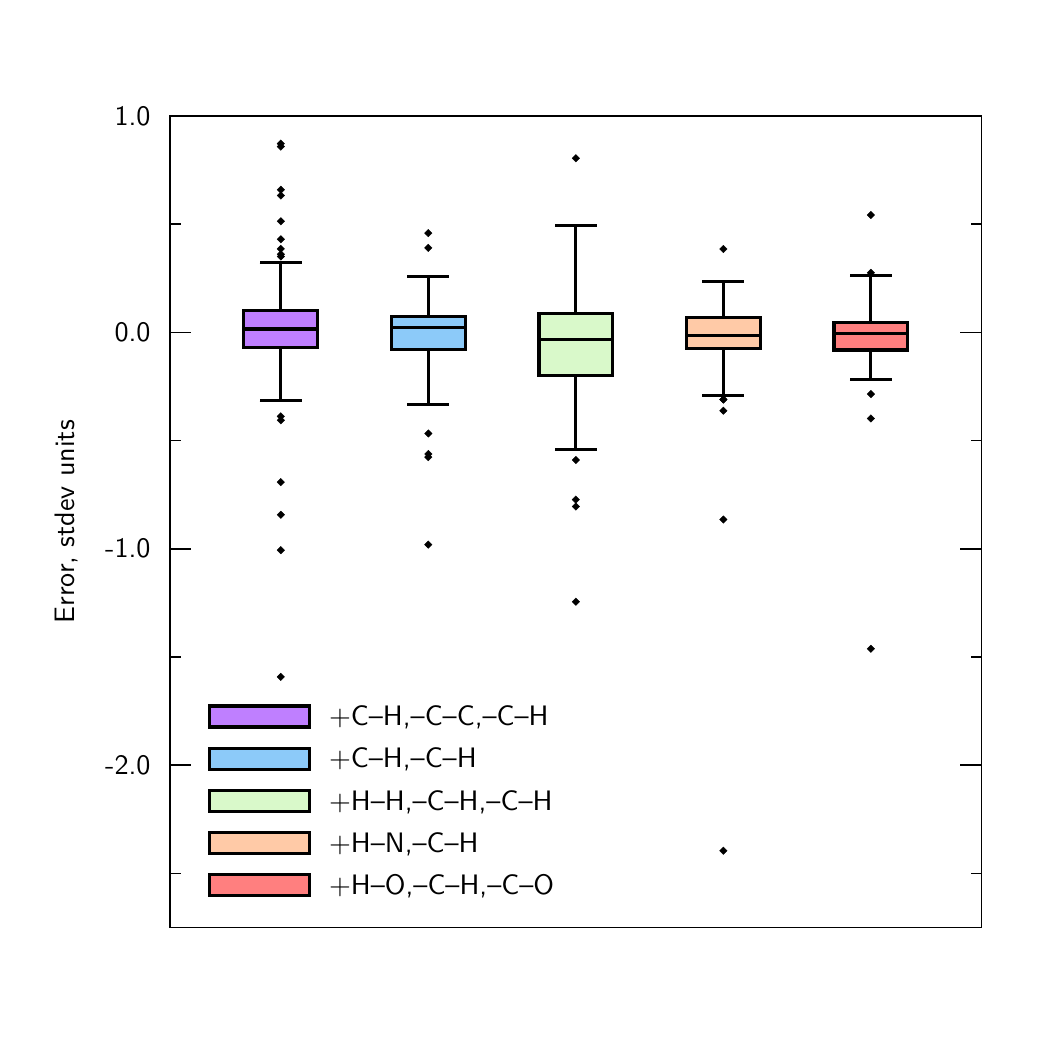}
\caption{
Box plots illustrating how \textsc{3DReact} (\textsc{InReact}$_M$ ``True'')
performs for the most common reaction types in the \gdb set,
when using explicit H nodes in the graphs.
The boxes range from the first to the third quartile of the datapoints.
The whiskers limit 90\% of the datapoints and the individual points illustrate outliers.
The points correspond to the test set of the first random split.
The errors are given in the target standard deviation (stdev) units (\SI{21.8}{kcal/mol}).
}
\label{fig:box_plot_with_H}
\end{figure}

Figure~\ref{fig:box_plot_with_H} illustrates the performance of \textsc{3DReact} ``True''
trained with explicit H-nodes. Compared to the model trained without explicit H-nodes
(Figure~\extref{fig:box_plot_reaction_classes} in the main text),
the error distribution in the different reaction types is more uniform,
since the \gdb dataset consists of many reactions involving breaking and forming \ce{H}--\ce{X} bonds,
which are better captured using a model with explicit H atoms.
Nevertheless, \textsc{3DReact} without explicit Hs already performs well
across the different reaction classes, in exchange for faster message passing.

\bigskip

Table \ref{tab:slatm-with-withoutH} shows the performance of SLATM$_d$+KRR with SLATM features constructed in three ways: with Hs (``with''), with H features removed after building the representation (``w/o after'') and H atoms removed before building the representation (``w/o before''). The first variation is the standard construction of SLATM features.\cite{amons2020}
The second variation is the closest possible version to ``implicit'' H regimes of \textsc{ChemProp} and \textsc{3DReact} since the H-only features (H--H and X--H--Y bins) are removed.
Nevertheless, other H-containing bins, \eg C--H or C--C--H, are still incorporated.
The last version removes H atoms completely from the system.
For the \gdb set, which is dependent on an accurate description of X--H bonds, errors increase systematically on the removal of H information. The same is seen to a lesser extent in the \cyclo set. For the \proparg, X--H bond changes have a minimal impact on the description of the reactions. SLATM$_d$ is run with explicit H atoms by default, since the representation is constructed from the xyz file directly without use of SMILES strings. Regardless of whether H atoms are in the SMILES strings, they are always present in the xyz file and therefore excluding H is nonsensical for SLATM.

\begin{table}[h!]
\begin{tabular}{@{}cccc@{}} \toprule
H mode
& \makecell{\gdb \\ ($\Delta E^\ddag$, kcal/mol)}
& \makecell{\cyclo \\ ($\Delta G^\ddag$, kcal/mol)}
& \makecell{\proparg \\ ($\Delta E^\ddag$, kcal/mol)}
\\ \midrule
with & $ 6.89 \pm 0.20 $ & $ 2.65 \pm 0.08 $ & $ 0.33 \pm 0.04 $\\
w/o after & $ 7.11 \pm 0.20 $ & $ 2.69 \pm 0.08 $ & $ 0.33 \pm 0.04 $\\
w/o before & $ 8.04 \pm 0.22 $ & $ 2.82 \pm 0.09 $ & $ 0.34 \pm 0.04 $\\
\bottomrule
\end{tabular}
\caption{
Performance of SLATM$_d$+KRR
with hydrogens in the representation
(H mode ``with'', as in the main text),
with hydrogens excluded after computing the representation
(H mode ``w/o after''),
and with hydrogens excluded before computing the representation
(H mode ``w/o before'').
MAEs are averaged over 10~folds of 80/10/10 splits (training/validation/test)
and reported together with standard deviations across folds.
}
\label{tab:slatm-with-withoutH}
\end{table}

%%%%%%%%%%%%%%%%%%%%%%%%%%%%%%%%%%%%%%%%%%%%%%%%%%%%%%%%%%%%%%%%%%%%%%%%%%%%%%%%%%%%%%%%%%%%%%%%%%%%
\clearpage
\section{Geometry sensitivity for the \cyclo dataset}
\label{sec:geom_cyclo}

\begin{figure}
\centering
\includegraphics[width=0.75\textwidth]{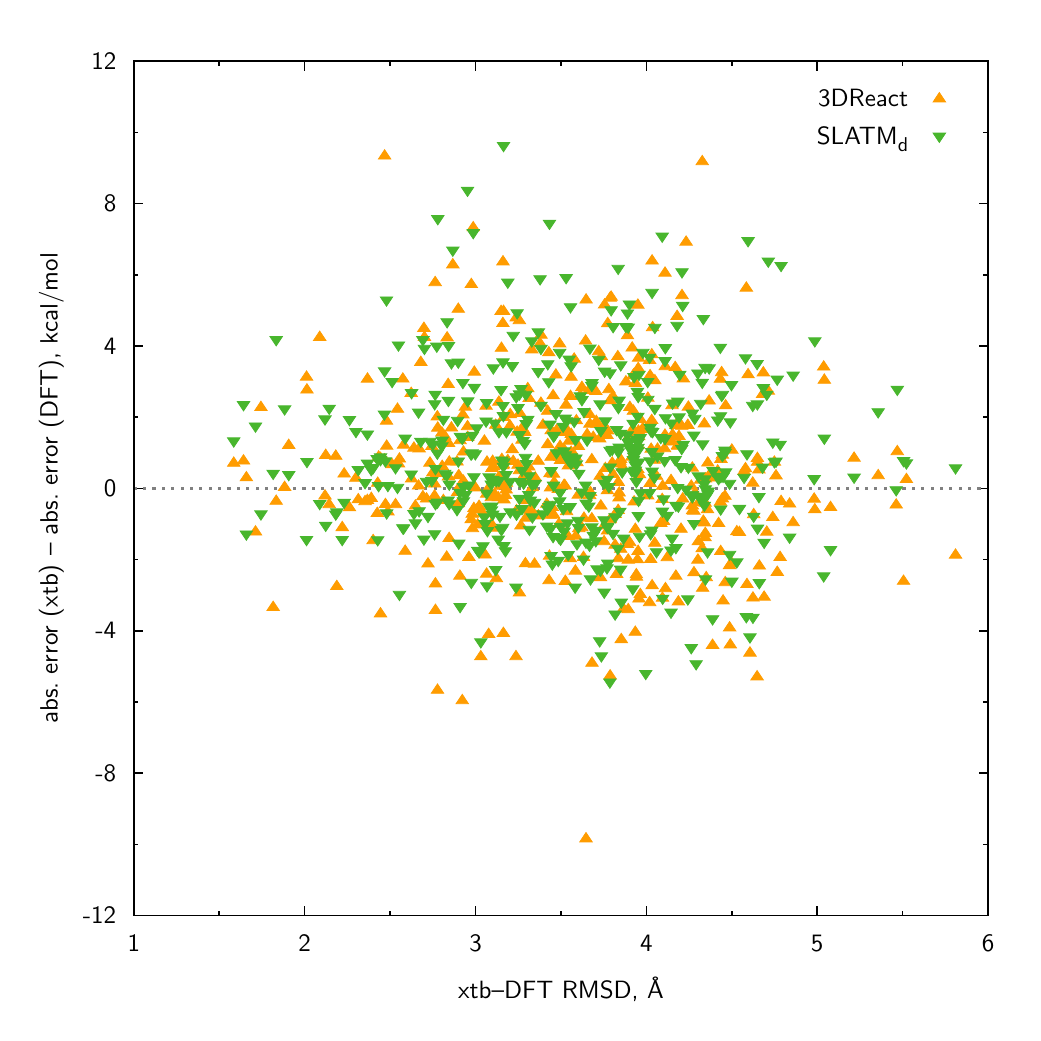}
\caption{
Difference between absolute prediction errors
using lower-quality GFN2-xTB\cite{gfn2_xtb} (xtb) and provided (DFT) geometries
\vs root-mean-square distance (RMSD) between said geometries
on the \cyclo set
for \textsc{3DReact} (\textsc{InReact}$_M$ ``True'') and SLATM$_d$+KRR.
}
\label{fig:cyclo_rmsd}
\end{figure}

Figure~\ref{fig:cyclo_rmsd} shows
the difference between absolute errors of models trained using GFN2-xTB (xtb) and DFT geometries
for both \textsc{3DReact} and SLATM$_d$+KRR
\vs
the root-mean-square distance (RMSD)
between the xtb and DFT geometries (as a measure of the agreement of the structures).
There is no noticeable trend for either model.
This is likely because a model trained on lower quality geometries
then struggles universally to predict barriers for lower quality geometries,
rather than resulting in larger errors for higher RMSD molecules.

RMSD\cite{Kabsch_1976} is computed as $\sqrt{\mathrm{RMSD}^2_{\mathrm{reactant}_1}+\mathrm{RMSD}^2_{\mathrm{reactant}_2}+\mathrm{RMSD}^2_\mathrm{product}}$
using the \texttt{rmsd}\cite{rmsd} python package.

\clearpage
\section*{References}
\bibliography{equireact.bib}